\documentclass{llncs}
\usepackage{amsmath,amssymb,amsfonts}
\usepackage{mathtools}
\usepackage[usenames,dvipsnames]{color}
\usepackage{xcolor}
\usepackage{xspace}
\usepackage{microtype}
\usepackage{todonotes}
\usepackage{colonequals}
\usepackage{wasysym}
 \usepackage{booktabs}

\usepackage{multirow}
\usepackage{multicol}

\usepackage{refcount}
\usepackage{paralist}

\usepackage{subfigure}
\usepackage{url}
\usepackage{appendix}
\usepackage{graphicx}
\usepackage{float}
\usepackage{algorithm}
\usepackage{listings}
\usepackage{newalg}
\usepackage{nicefrac}
\usepackage{tikz} 
\usepackage{pgfplots}
\usepackage{wrapfig}
\usepackage{hyperref}

\usepackage{footnote}
\usepackage{tablefootnote}
\usepackage{array}

\usetikzlibrary{arrows,decorations.pathmorphing,positioning,fit,trees,shapes,shadows,automata,calc} 

\usepackage{pdfpages}

\usepackage{macros} 

\begin{document}
  \title{Parameter Synthesis for Markov Models:\\ Faster Than Ever}


\author{Omitted for Review\\}

\author{Tim Quatmann\inst{1} \and Christian Dehnert\inst{1} \and Nils Jansen\inst{2} \and\\  Sebastian Junges\inst{1} \and Joost-Pieter Katoen\inst{1}}

\institute{RWTH Aachen University \and University of Texas at Austin}
\maketitle

  \begin{abstract}

We propose a conceptually simple technique for verifying probabilistic models whose transition probabilities are parametric. The key is to replace parametric transitions by nondeterministic choices of extremal values. Analysing the resulting parameter-free model using off-the-shelf means yields (refinable) lower and upper bounds on probabilities of regions in the parameter space. The technique outperforms the existing analysis of parametric Markov chains by several orders of magnitude regarding both run-time and scalability. Its beauty is its applicability to various probabilistic models. It in particular provides the first sound and feasible method for performing parameter synthesis of Markov decision processes.

\end{abstract}

%
  \section{Introduction}
\label{sec:introduction}

The key procedure in probabilistic model checking is computing reachability probabilities: What is the probability to reach some target state?
For models exhibiting nondeterminism, such as Markov decision processes (MDPs), the probability to reach a state is subject to resolving the nondeterminism, and one considers minimal and maximal reachability probabilities.
Model checkers support these procedures, \eg, \prism~\cite{KNP11} and \tool{iscasMc}~\cite{iscasmc}.
Successful applications to models of hundreds of millions of states have been reported, and extensions to stochastic games exist~\cite{DBLP:conf/tacas/ChenFKPS13}.

This paper treats \emph{parameter synthesis} in Markov models.
Given a model whose transition probabilities are (polynomials over) variables, and a reachability specification---e.g., the likelihood to reach a bad state should be below 10$^{-6}$---the parameter synthesis problem aims at finding all parameter values for which the parametric model satisfies the specification.
In practise, this amounts to partition the parameter space into \emph{safe} and \emph{unsafe} regions with a large (say, $> 95\%$) coverage.
For a system in which components are subject to random failures, parameter synthesis is thus able to obtain the maximal tolerable failure probability of the components while ensuring the system's specification.

Parametric probabilistic models have various applications as witnessed by several recent works.
Model repair~\cite{bartocci2011model} exploits parametric Markov chains (MCs) to tune the parameters of the model.
In quality-of-service analysis of software, parameters are used to model the unquantified estimation errors in log data~\cite{calinescu_ieee_tr_2016}.
Ceska \emph{et al.}~\cite{DBLP:conf/cmsb/CeskaDKP14} consider the problem of synthesising rate parameters in stochastic biochemical networks.
Parametric probabilistic models are also used to rank patches in the repair of software~\cite{DBLP:conf/popl/LongR16} and for computing perturbation bounds~\cite{rosenblum-et-al-atva-2014,su-et-al-icse-2016-qosevaluation}. 
The main problem though is that current parametric probabilistic model-checking algorithms cannot cope with the complexity of these applications.
Their scalability is restricted to a couple of thousands of states and a few (preferably independent) parameters, and models with nondeterminism are out of reach.
(The only existing algorithm~\cite{DBLP:conf/nfm/HahnHZ11} for parametric MDPs uses an unsound heuristic in its implementation to improve scalability.)

We present an algorithm that overcomes all these limitations: It is scalable to millions of states, several (dependent) parameters, and---perhaps most importantly---provides the first sound and feasible technique to do parameter synthesis of parametric MDPs.

The key technique used so far is computing a rational function (in terms of the parameters) expressing the reachability probability in a parametric MC.
Tools like \tool{PARAM}~\cite{PARAM10}, \tool{PRISM}~\cite{KNP11}, and \prophesy~\cite{dehnert-et-al-cav-2015} exploit (variants of) the state elimination approach by Daws~\cite{Daw04} to obtain such a function which conceptually allows for many types of analysis.
While state elimination is feasible for millions of states~\cite{dehnert-et-al-cav-2015}, it does not scale well in the number of different parameters.
Moreover, the size of the obtained functions often limits the practicability as analysing the (potentially large) rational function via SMT solving~\cite{dehnert-et-al-cav-2015} is often not feasible.



\begin{figure}[t]
	\centering
	\subfigure[Parametric model]{
		\scalebox{0.85}{
			\begin{tikzpicture}[scale=1, nodestyle/.style={draw,circle},baseline=(s0)]
    
    \node [nodestyle] (s0) at (0,0) {$s_0$};
    \node [nodestyle] (s1) [on grid, right= 1.5cm of s0] {$s_1$};
    \node [nodestyle] (s3) [on grid, below right=1.5cm and 0.75cm of s0, yshift=0.3cm] {$s_3$};
    \node [nodestyle, accepting] (s2) [on grid, right=1.5cm of s1] {$s_2$};
    
    \draw ($(s0)+(0,0.7)$) edge[->] (s0);
    \draw (s0) edge[->] node[above] {\scriptsize$x$} (s1);
    \draw (s0) edge[->] node[left] {\scriptsize$1{-}x$} (s3);

    \draw (s1) edge[->] node[above] {\scriptsize$1{-}x$} (s2);
    \draw (s1) edge[->] node[right] {\scriptsize$x$} (s3);

    \draw (s3) edge[loop right, ->] node[auto] {\scriptsize$1$} (s3);
    \draw (s2) edge[loop below, ->] node[auto] {\scriptsize$1$} (s2);
    
\end{tikzpicture}
		}
		\label{fig:introductionExample:pMC}
	}
	\subfigure[Relaxation]{
		\scalebox{0.85}{
			\begin{tikzpicture}[scale=1, nodestyle/.style={draw,circle},baseline=(s0)]
    
    \node [nodestyle] (s0) at (0,0) {$s_0$};
    \node [nodestyle] (s1) [on grid, right=1.5cm of s0] {$s_1$};
    \node [nodestyle] (s3) [on grid, below right=1.5cm and 0.75cm of s0, yshift=0.3cm] {$s_3$};
    \node [nodestyle, accepting] (s2) [on grid, right=1.5cm of s1] {$s_2$};
    
    \draw ($(s0)+(0,0.7)$) edge[->] (s0);
    \draw (s0) edge[->] node[above] {\scalebox{0.8}{$ x^{s_{0}}$}} (s1);
    \draw (s0) edge[->] node[left] {\scalebox{0.8}{$1{-}x^{s_0}$}} (s3);

    \draw (s1) edge[->] node[above] {\scalebox{0.8}{$1{-}x^{s_1}$}} (s2);
    \draw (s1) edge[->] node[right] {\scalebox{0.8}{$x^{s_1}$}} (s3);

    \draw (s3) edge[loop right, ->] node[auto] {\scalebox{0.8}{$1$}} (s3);
    \draw (s2) edge[loop below, ->] node[auto] {\scalebox{0.8}{$1$}} (s2);
    
\end{tikzpicture}
		}
		\label{fig:introductionExample:rel}
	}
	\subfigure[Substitution]{
		\scalebox{0.85}{
			\begin{tikzpicture}[scale=1, nodestyle/.style={draw,circle},baseline=(s0)]

\node [nodestyle] (s0) at (0,0) {$s_0$};
\node [nodestyle] (s1) [on grid, right= 2cm of s0] {$s_1$};
\node [nodestyle] (s3) [on grid, below right=1.5cm and 1cm of s0, yshift=0.3cm] {$s_3$};
\node [nodestyle, accepting] (s2) [on grid, right=1.5cm of s1] {$s_2$};

\draw ($(s0)+(0,0.7)$) edge[->] (s0);

\draw (s0) edge[->, dashed, bend right=15] node[above] {\scriptsize$0.3$} (s1);
\draw (s0) edge[->, bend left] node[above] {\scriptsize$0.6$} (s1);

\draw (s0) edge[->, dashed, bend left=0] node[right, pos=0.33] {\scriptsize$0.7$} (s3);
\draw (s0) edge[->, bend right] node[left] {\scriptsize$0.4$} (s3);

\draw (s1) edge[->, dashed, bend right] node[above] {\scriptsize$0.7$} (s2);
\draw (s1) edge[->, bend left] node[above] {\scriptsize$0.4$} (s2);

\draw (s1) edge[->, dashed, bend right=0] node[left, pos=0.33] {\scriptsize$0.3$} (s3);
\draw (s1) edge[->,  bend left] node[right] {\scriptsize$0.6$} (s3);

\draw (s3) edge[loop right, ->] node[auto] {\scriptsize$1$} (s3);
\draw (s2) edge[loop below, ->] node[auto] {\scriptsize$1$} (s2);
\end{tikzpicture}
		}
		\label{fig:introductionExample:approx}
	}
       \vspace*{-0.3cm}
	\caption{Two biased coin tosses and the specification ``First \emph{heads} then \emph{tails}''.}
	\vspace{-4mm}
	\label{fig:introductionExample}
\end{figure}
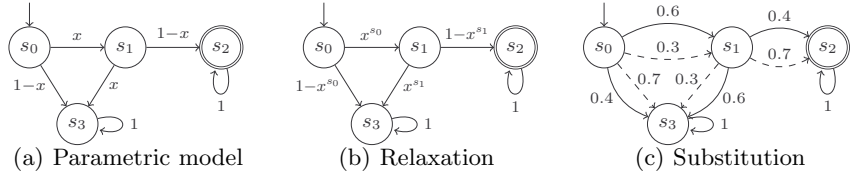
This paper takes a completely different approach: \emph{Parameter lifting}.
Consider the parametric MC in Fig.~\ref{fig:introductionExample:pMC} modelling two subsequent tosses of a biased coin, where the probability for \emph{heads} is $x$.
Inspired by an observation made in \cite{BCDS13} on continuous time Markov chains, we first equip each state with a fresh parameter, thus removing parameter dependencies; the outcome (referred to as \emph{relaxation}) is depicted in
Fig.~\ref{fig:introductionExample:rel}.
Now, for each function over these state parameters, we compute \emph{extremal values}, \ie, maximal and minimal probabilities. 
The key idea is to replace the (parametric) probabilistic choice at each state by a \emph{nondeterministic choice} between these extremal values; we call this \emph{substitution}.
This is exemplified in Fig.~\ref{fig:introductionExample:approx}, assuming \emph{heads} has a likelihood in $[0.3,0.6]$.
The resulting (non-parametric) model can be verified using off-the-shelf, efficient algorithms.
Applying this procedure to a parametric MC (as in the example) yields a parameter-free MDP.
Parameter lifting thus boils down to verify an MDP and avoids computing rational functions and SMT solving. 
The beauty of this technique is that it can be applied to parametric MDPs without much further ado.
Parameter lifting of a parametric MDP yields a parameter-free two-player stochastic game (SG).
SGs and MDPs can be solved using techniques such as value and policy iteration. Note that the theoretical complexity for solving MDPs is lower than for SGs.

This paper presents the details of parameter lifting, and proves the correctness for parametric Markov models whose parameters are given as \emph{multi-affine polynomials}.
This covers a rich class of models, \eg, all parametric benchmarks available at the \tool{PARAM} webpage are of this form.
Experiments demonstrate the feasibility: The parameter lifting approach can treat Markov models of millions of states with thousands of parametric transitions.
This applies to parametric MCs as well as MDPs.
Parameter lifting achieves a parameter space coverage of at least 95$\%$ rather quickly.
This is out of reach for competitive techniques such as SMT-based~\cite{dehnert-et-al-cav-2015} and sampling-based~\cite{DBLP:conf/nfm/HahnHZ11}  parameter synthesis.

  \section{Preliminaries}
\label{sec:preliminaries}
%
%
Let $\Var$ be a finite set of \emph{parameters} over the domain $\R$ ranged over by $x, y, z$.
A \emph{valuation} for $\Var$ is a function $u \colon \Var \to \R$.
Let $\functions[\Var]$ denote the set of \emph{multi-affine multivariate polynomials} $\pol$ over $\Var$ satisfying $\pol = \sum_{i \le m} a_i \cdot \prod_{x \in \Var_i} x$ for suitable $m \in \N$, $a_i \in \Q$, and $\Var_i \subseteq \Var$ (for $i \le m$).
$\functions[\Var]$ does not contain polynomials where a variable has a degree greater than $1$, \eg, $x\cdot y \in \functions[\Var]$ but $x^2  \notin \functions[\Var]$.
We write $\pol=0$ if $\pol$ can be reduced to $0$, and $\pol\not=0$ otherwise. 
%
%
%
Applying the valuation $u$ to $\pol \in \functions[\Var]$ results in a real number $\pol[u]\in\R$, obtained from $\pol$ by replacing each occurrence of variable $x$ in $\pol$ by $u(x)$.

\subsection{Probabilistic Models}\label{sec:preliminaries:models}
We consider different types of parametric (discrete) probabilistic models.
They can all be seen as transition systems (with a possible partition of the state space into two sets) where the transitions are labeled with polynomials in $\functions[\Var]$.

\begin{definition}[Parametric probabilistic models]\label{def:parametric_model}
A \emph{parametric stochastic game (pSG)} is a tuple $\psgInitGeneric$ with a finite set $S$ of states such that $S = \spOne\uplus\spTwo$, a finite set $\Var$ of parameters over $\R$, an initial state $\sinit \in S$, a finite set $\Act$ of actions, and a transition function $\probmdp \colon S \times \Act \times S \rightarrow \functions[\Var]$ satisfying: 
$
\Act(s) \neq \emptyset \mbox{ whith } \Act(s) = \{\act \in \Act \mid \exists s'\in S.\,\probmdp(s,\act,s') \neq 0\}.
$
\begin{compactitem}
\item
$\psgGeneric$ is a \emph{parametric Markov decision process (pMDP)} if $\spOne=\emptyset$ or $\spTwo=\emptyset$.
\item
pMDP $\psgGeneric$ is a \emph{parametric Markov chain (pMC)} if $|\Act(s)|=1$ for all $s \in S$.
\end{compactitem}
\end{definition}
We will refer to pMCs by $\dtmc$, to pMDPs by $\mdp$ and to pSGs by $\psg$.
pSGs are two-player parametric stochastic games involving players $\pOne$ and $\pTwo$ with states in $\spOne$ and $\spTwo$, respectively, whose transition probabilities are represented by polynomials from $\functions[\Var]$.
The players \emph{nondeterministically} choose an action at each state and the successors are intended to be determined \emph{probabilistically} as defined by the transition function.
$\Act(s)$ is the set of \emph{enabled} actions at state $s$.
As $\Act(s)$ is non-empty for all $s \in S$, there are no deadlock states.
For state $s$ and action $\alpha$, we set $\Var_s^\alpha = \{x \in \Var \mid x \text{ occurs in } \probmdp(s,\alpha,s') \text{ for some } s'\in S \}$.

\noindent pMDPs and pMCs are one- and zero-player parametric stochastic games, respectively.
As pMCs have in fact just a single enabled action at each state, we omit this action in the notation and just write $\probmdp(s,s')$ and $V_s$.
	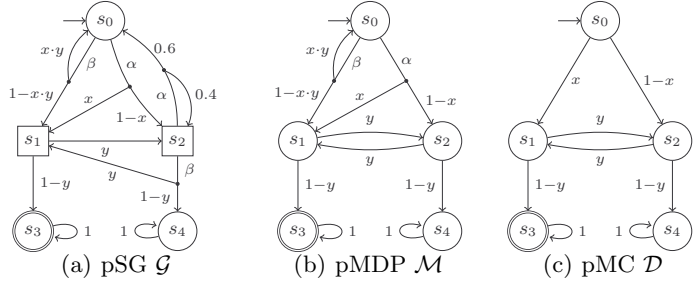
\begin{figure}[t]
		\centering
		\subfigure[pSG $\psg$]{
			\scalebox{\picscale}{
				\begin{tikzpicture}[scale=1, nodestyle/.style={draw,circle},baseline=(s0)]
    
    \node [nodestyle] (s0) at (0,0) {$s_0$};
    \node [nodestyle, rectangle, minimum height=14pt] (s1) [on grid, below left=2cm and 1.2cm of s0] {$s_1$};
    \node [nodestyle, rectangle, minimum height=14pt] (s2) [on grid, below right=2cm and 1.2cm of s0] {$s_2$};
    \node [nodestyle, accepting] (s3) [on grid, below=\distsonesthree of s1] {$s_3$};
    \node [nodestyle] (s4) [on grid, below=\distsonesthree of s2] {$s_4$};
    
    \draw ($(s0)-(0.7,0)$) edge[->] (s0);
    \draw (s0) edge[->] node[pos=0.3, right] {\scriptsize$\beta$} node[draw, circle,  inner sep=0.5pt, fill] (s0beta) {} node [pos=0.65, left] {\scriptsize$1{-}x{\cdot}y$} (s1);
    \draw (s0beta) edge[->, bend left=30] node[left] {\scriptsize$x{\cdot} y$} (s0);
    \draw (s0) edge[->, bend right=15] node[pos=0.25, right] {\scriptsize$\alpha$} node[draw, circle,  inner sep=0.5pt, fill] (s0alpha) {} node [pos=0.9, left] {\scriptsize$1{-}x$} (s2);
    \draw (s0alpha) edge[->] node[above] {\scriptsize$x$} (s1);

     \draw (s1) edge[->] node[below] {\scriptsize$y$} (s2);
     \draw (s1) edge[->] node[right] {\scriptsize$1{-}y$} (s3);
    
    \draw (s2) edge[->, bend right] node[pos=0.25, left] {\scriptsize$\alpha$} node[draw, circle,  inner sep=0.5pt, fill] (s2alpha) {} node [pos=0.75, right] {\scriptsize$0.6$} (s0);
     \draw (s2alpha) edge[bend left=45, ->] node[right] {\scriptsize$0.4$} (s2);
    \draw (s2) edge[->] node[pos=0.25, right] {\scriptsize$\beta$} node[draw, circle,  inner sep=0.5pt, fill] (s2beta) {} node [pos=0.75, left] {\scriptsize$1{-}y$} (s4);
     \draw (s2beta) edge[->] node[below] {\scriptsize$y$} (s1);
    
    \draw (s3) edge[loop right, ->] node[auto] {\scriptsize$1$} (s3);
    
    \draw (s4) edge[loop left, ->] node[auto] {\scriptsize$1$} (s3);

\end{tikzpicture}
			}
			\label{fig:models:psg}
		}
		\subfigure[pMDP $\pmdp$]{ 
			\scalebox{\picscale}{
				\begin{tikzpicture}[scale=1, nodestyle/.style={draw,circle},baseline=(s0)]
    
    \node [nodestyle] (s0) at (0,0) {$s_0$};
    \node [nodestyle] (s1) [on grid, below left=2cm and 1.2cm of s0] {$s_1$};
    \node [nodestyle] (s2) [on grid, below right=2cm and 1.2cm of s0] {$s_2$};
    \node [nodestyle, accepting] (s3) [on grid, below=\distsonesthree of s1] {$s_3$};
    \node [nodestyle] (s4) [on grid, below=\distsonesthree of s2] {$s_4$};
    
    \draw ($(s0)-(0.7,0)$) edge[->] (s0);
    \draw (s0) edge[->] node[pos=0.3, right] {\scriptsize$\beta$} node[draw, circle,  inner sep=0.5pt, fill] (s0beta) {} node [pos=0.65, left] {\scriptsize$1{-}x{\cdot}y$} (s1);
    \draw (s0beta) edge[->, bend left=30] node[left] {\scriptsize$x{\cdot} y$} (s0);
    \draw (s0) edge[->, bend right=0] node[pos=0.25, right] {\scriptsize$\alpha$} node[draw, circle,  inner sep=0.5pt, fill] (s0alpha) {} node [pos=0.75, right] {\scriptsize$1{-}x$} (s2);
    \draw (s0alpha) edge[->] node[above] {\scriptsize$x$} (s1);
    
     \draw (s1) edge[bend left=10, ->] node[above] {\scriptsize$y$} (s2);
     \draw (s1) edge[->] node[right] {\scriptsize$1{-}y$} (s3);
     
     \draw (s2) edge[bend left=10, ->] node[below] {\scriptsize$y$} (s1);
     \draw (s2) edge[->] node[left] {\scriptsize$1{-}y$} (s4);
    
    \draw (s3) edge[loop right, ->] node[auto] {\scriptsize$1$} (s3);
    
    \draw (s4) edge[loop left, ->] node[auto] {\scriptsize$1$} (s3);

\end{tikzpicture}
			}
			\label{fig:models:pmdp}
		}
		\subfigure[pMC $\pdtmc$]{ 
			\scalebox{\picscale}{
				\begin{tikzpicture}[scale=1, nodestyle/.style={draw,circle},baseline=(s0)]
    
    \node [nodestyle] (s0) at (0,0) {$s_0$};
    \node [nodestyle] (s1) [on grid, below left=2cm and 1.2cm of s0] {$s_1$};
    \node [nodestyle] (s2) [on grid, below right=2cm and 1.2cm of s0] {$s_2$};
    \node [nodestyle, accepting] (s3) [on grid, below=\distsonesthree of s1] {$s_3$};
    \node [nodestyle] (s4) [on grid, below=\distsonesthree of s2] {$s_4$};
    
    \draw ($(s0)-(0.7,0)$) edge[->] (s0);
    \draw (s0) edge[->] node[right] {\scriptsize$x$} (s1);
    \draw (s0) edge[->] node[right] {\scriptsize$1{-}x$} (s2);

     \draw (s1) edge[bend left=10, ->] node[above] {\scriptsize$y$} (s2);
     \draw (s1) edge[->] node[right] {\scriptsize$1{-}y$} (s3);
    
     \draw (s2) edge[bend left=10, ->] node[below] {\scriptsize$y$} (s1);
     \draw (s2) edge[->] node[left] {\scriptsize$1{-}y$} (s4);
    
    \draw (s3) edge[loop right, ->] node[auto] {\scriptsize$1$} (s3);
    
    \draw (s4) edge[loop left, ->] node[auto] {\scriptsize$1$} (s3);

\end{tikzpicture}
			}
			\label{fig:models:pmc}
		}
		\vspace{-2mm}
		\caption{The considered types of parametric probabilistic models.}
		\label{fig:models}
		\vspace{-4mm}
	\end{figure}
\begin{example}
Fig.~\ref{fig:models} depicts (a.) a pSG, (b.) a pMDP, and (c.) a pMC with parameters $\{x,y\}$.
	The states of the players $\pOne$ and $\pTwo$ are depicted with circles and rectangles, respectively.
	The initial state is indicated by an arrow; target states have double lines.
	We draw a transition from state $s$ to $s'$ and label it with $\alpha$ and $\probmdp(s,\alpha,s')$ whenever $\probmdp(s,\alpha,s')\neq 0$.
    If $|\Act(s)|=1$, the action is omitted.
\end{example}
\begin{remark}
In the literature~\cite{param_sttt,dehnert-et-al-cav-2015}, the images of transition functions (of pMCs) are rational functions, \ie, fractions of polynomials.
This is mainly motivated by the usage of state elimination for computing functions expressing reachability probabilities.
As our approach does not rely on state elimination, the set of considered functions can be simplified.
The restriction to polynomials in $\functions[\Var]$ is realistic; \emph{all} benchmarks from the \param webpage~\cite{param_website} are of this form. We will exploit this restriction in our proof of Theorem~\ref{lem:optimalValuation}.
\end{remark}
\begin{definition}[Stochastic game]
A pSG $\sg$ is a \emph{stochastic game (SG)} if $\probmdp \colon S \times \Act \times S \rightarrow [0,1]$ and $\sum_{s'\in S}\probmdp(s,\act,s') = 1$ for all $s \in S \mbox{ and } \act \in \Act(s)$.
\end{definition}
Analogously, MCs and MDPs are defined as special cases of pMCs and pMDPs.
Thus, a model is \emph{parameter-free} if all transition probabilities are constant. 

\medskip\noindent\emph{Valuations and rewards.}
Applying a \emph{valuation} $u$ to parametric model $\psgGeneric$, denoted $\psgGeneric$$[u]$, replaces each polynomial $f$ in $\psgGeneric$ by $f[u]$.
We call $\psgGeneric$$[u]$ the \emph{instantiation} of $\psgGeneric$ at $u$.
The typical application of $u$ is to replace the transition function $f$ by the probability $f[u]$. 
A valuation $u$ is \emph{well-defined} for $\psgGeneric$ if the replacement yields probability distributions, \ie, if $\psgGeneric$$[u]$ is an MC, an MDP, or an SG, respectively.

Parametric probabilistic models are extended with \emph{rewards} (or dually, costs) by adding a \emph{reward function} $\rewFunction \colon S \rightarrow \functions[\Var]$ which assigns rewards to states of the model.
Intuitively, the reward $\rewFunction(s)$ is earned upon leaving the state $s$.

\medskip\noindent\emph{Schedulers.}
The nondeterministic choices of actions in pSGs and pMDPs can be resolved using \emph{schedulers}\footnote{Also referred to as adversaries, strategies, or policies.}.
In our setting it suffices to consider memoryless deterministic schedulers~\cite{Var85}. 
For more general definitions we refer to~\cite{BK08}.
\begin{definition}{\bf (Scheduler)}\label{def:scheduler}
	A \emph{scheduler} for pMDP $\pmdpInit$ is a function $\sched\colon S\rightarrow\Act$ with $\sched(s)\in \Act(s)$ for all $s\in S$.  
\end{definition}
Let $\Sched{\mdp}$ denote the set of all schedulers for $\pmdp$.
Applying a scheduler to a pMDP yields an \emph{induced parametric Markov chain}, as all nondeterminism is resolved, \ie, the transition probabilities are obtained \wrt the choice of actions.
\begin{definition}{\bf (Induced pMC)}\label{def:induced_dtmc} 
	For pMDP $\pmdpInit$ and scheduler $\sched\in\Sched{\pmdp}$, the \emph{pMC induced by $\pmdp$ and $\sched$} is $\pmdp^\sched=(S, \Var, \sinit,\probmdp^\sched)$ where
	\begin{align*}
		\probmdp^\sched(s,s')= \probmdp(s,\sched(s),s') \quad \mbox{ for all } s,s'\in S\ .
	\end{align*} 
\end{definition}
%
\noindent Resolving nondeterminism in an SG requires to have individual schedulers for each player. 
For $\spOne$ and $\spTwo$ we need schedulers $\sched \in \Sched[\pOne]{\sg}$ and $\altsched \in \Sched[\pTwo]{\sg}$ of the form $\sched \colon {\spOne}\rightarrow\Act$ and $\altsched\colon \spTwo\rightarrow\Act$. The induced pMC $\sg^{\sched, \altsched}$ of a pSG $\sg$ with schedulers $\sched$ and $\altsched$ for both players is defined analogously to the one for pMDPs.
\begin{example}
	\label{ex:inducedModels}
	Reconsider the models $\psg$, $\pmdp$, and $\pdtmc$ as shown in Fig.~\ref{fig:models}.
	For schedulers $\sched$, $\altsched$ with  $\sched(s_2)=\beta$ and $\altsched(s_0) = \alpha$, the induced pMCs satisfy $\psg^{\sched, \altsched} = \pmdp^\altsched = \pdtmc$.
\end{example}

\subsection{Properties of Interest}
As specifications we consider \emph{reachability properties} and \emph{expected reward properties}.
We first define these properties on MCs and then discuss the other models.

\medskip\noindent\emph{Properties on MCs.}
For MC $\dtmc$ with state space $S$, let $\reachPrs{\dtmc}{s}{T}$ denote the probability to reach a set of target states $T \subseteq S$ from  state $s\in S$ within $\pdtmc$; simply $\reachPrT[\dtmc]$ refers to this specific probability for the initial state $\sinit$.
We use a standard probability measure on infinite paths through an MC as defined in~\cite[Ch.\ 10]{BK08}.
For threshold $\lambda\in [0,1]$, the \emph{reachability property} asserting that a target state is to be reached with probability at most $\lambda$ is denoted $\reachPropSymbol = \reachProplT$.
The property is satisfied by $\dtmc$, written $\dtmc \models \reachPropSymbol$, iff $\reachPrT[\dtmc]\leq\lambda$.
(Comparisons like $<$, $>$, and $\geq$ are treated in a similar way.)

The reward of a path through an MC $\dtmc$ until $T$ is the sum of the rewards of the states visited along on the path before reaching $T$.
The expected reward of a finite path is given by its probability times its reward.
Given $\reachPrT[\dtmc] = 1$, the expected reward of reaching $T \subseteq S$, is the sum of the expected rewards of all paths to reach $T$.
An expected reward property is satisfied if the expected reward of reaching $T$ is bounded by a threshold $\kappa \in \R$.
Formal definitions can be found in e.g., ~\cite[Ch.\ 10]{BK08}.

\medskip\noindent\emph{Properties on nondeterministic models.}
In order to define a probability measure for MDPs and SGs, the nondeterminism has to be resolved.
A reachability property $\reachProplT$ is satisfied for an MDP $\mdp$ iff it holds for all induced MCs:
\begin{align*}
	\mdp \models \reachProplT \iff
	\big(\max_{\sched \in \Sched{\mdp}} \reachPrT[\mdp^\sched] \big) \leq \lambda \text{.}
\end{align*}
Satisfaction of a property $\varphi$ for an SG $\sg$ depends on the objectives of both players.
We write $\sg \models_{\triangle} \varphi$ iff players in $\triangle \subseteq \{\pOne, \pTwo\}$ can enforce that $\varphi$ holds, \eg,
\begin{align*}
\sg \models_{\{\pOne\}} \reachProplT \iff
\big( \min_{\sched \in \Sched[\pOne]{\sg}}  \max_{\altsched \in \Sched[\pTwo]{\sg}} \reachPrT[\sg^{\sched, \altsched}] \big) \leq \lambda \text{.}
\end{align*}
Computing the maximal (or minimal) probability to reach a set of target states from the initial state can be done using standard techniques, such as linear programming, value iteration or policy iteration~\cite{Put94}.

The satisfaction relation for expected reward properties is defined analogously.
As usual, we write $\psgGeneric \models \neg \varphi$ whenever $\psgGeneric \not\models \varphi$.

  \section{Regional Model Checking of Markov Chains}
\label{sec:regionsMC}
In the following, we consider sets of valuations that map each parameter to a value within a given interval. 
We present an approximative approach to check all instantiations of a pMC with respect to a valuation in such a set.
This consists of three steps: Formalising regions and the considered problem, construction of the sound over-approximation, and reduction to an MDP problem.
%
%
\subsection{Regions}
\label{sec:regionsMC:regions}

\begin{definition}{\bf (Region)}\label{def:region}
Given a set of parameters $V = \{x_1, \hdots x_n\}$ and rational \emph{parameter bounds} $B(x_i) = \{b_1, b_2\}$. The parameter bounds induce a \emph{parameter interval} $I(x_i) = [b_1, b_2]$ with $b_1 \leq b_2$.  
The set of valuations $\{ u \mid \forall x_i \in V.\,u(x_i) \in I(x_i) \}$ is called a \emph{region (for $V$)}.  
\end{definition}
The regions we consider correspond to $\bigtimes_{x \in V} I(x)$, \ie, they are \emph{hyperrectangles}.

We aim to identify sets of instantiated models by regions. That is, regions represent instantiations $\psgGeneric[u]$ of a parametric model $\psgGeneric$. As these instantiations are only well-defined under some restrictions, we lift these restrictions to regions.

\begin{figure}[t]
\vspace{-5mm}
  \centering
  \subfigure[$\pdtmc$]{
  \centering
    \scalebox{\picscale}{
      \begin{tikzpicture}[scale=1, nodestyle/.style={draw,circle},baseline=(s0)]
    
    \node [nodestyle] (s0) at (0,0) {$s_0$};
    \node [] (leftdummy)  [on grid, left=1.2cm of s0] {};
    \node [] (rightdummy) [on grid, right=1.2cm of s0] {};
    \node [nodestyle] (s1) [on grid, below=\distforsone of leftdummy] {$s_1$};
    \node [nodestyle] (s2) [on grid, below=\distforsone of rightdummy] {$s_2$};
    \node [nodestyle, accepting] (s3) [on grid, below=\distsonesthree of s1] {$s_3$};
    \node [nodestyle] (s4) [on grid, below=\distsonesthree of s2] {$s_4$};
    
    \draw ($(s0)-(0.7,0)$) edge[->] (s0);
    \draw (s0) edge[->] node[right] {\scriptsize$x$} (s1);
    \draw (s0) edge[->] node[right] {\scriptsize$1{-}x$} (s2);

     \draw (s1) edge[bend left=15, ->] node[above] {\scriptsize$y$} (s2);
     \draw (s1) edge[->] node[right] {\scriptsize$1{-}y$} (s3);
    
     \draw (s2) edge[bend left=15, ->] node[below] {\scriptsize$y$} (s1);
     \draw (s2) edge[->] node[left] {\scriptsize$1{-}y$} (s4);
    
    \draw (s3) edge[loop right, ->] node[auto] {\scriptsize$1$} (s3);
    
    \draw (s4) edge[loop left, ->] node[auto] {\scriptsize$1$} (s3);

\end{tikzpicture}
    }
    \label{fig:pmc:D}
  }
  \subfigure[{$\pdtmc[u]$}]{
  \centering
    \scalebox{\picscale}{
      \begin{tikzpicture}[scale=1, nodestyle/.style={draw,circle},baseline=(s0)]
    
    \node [nodestyle] (s0) at (0,0) {$s_0$};
    \node [] (leftdummy)  [on grid, left=1.2cm of s0] {};
    \node [] (rightdummy) [on grid, right=1.2cm of s0] {};
    \node [nodestyle] (s1) [on grid, below=\distforsone of leftdummy] {$s_1$};
    \node [nodestyle] (s2) [on grid, below=\distforsone of rightdummy] {$s_2$};
    \node [nodestyle, accepting] (s3) [on grid, below=\distsonesthree of s1] {$s_3$};
    \node [nodestyle] (s4) [on grid, below=\distsonesthree of s2] {$s_4$};
    
    \draw ($(s0)-(0.7,0)$) edge[->] (s0);
    \draw (s0) edge[->] node[right] {\scriptsize$0.8$} (s1);
    \draw (s0) edge[->] node[right] {\scriptsize$0.2$} (s2);

     \draw (s1) edge[bend left=15, ->] node[above] {\scriptsize$0.6$} (s2);
     \draw (s1) edge[->] node[right] {\scriptsize$0.4$} (s3);
    
     \draw (s2) edge[bend left=15, ->] node[below] {\scriptsize$0.6$} (s1);
     \draw (s2) edge[->] node[left] {\scriptsize$0.4$} (s4);
    
    \draw (s3) edge[loop right, ->] node[auto] {\scriptsize$1$} (s3);
    
    \draw (s4) edge[loop left, ->] node[auto] {\scriptsize$1$} (s3);

\end{tikzpicture}
    }
    \label{fig:pmc:Du}
  }
  \subfigure[$\rel{\pdtmc}$]{
  \centering
    \scalebox{\picscale}{
      \begin{tikzpicture}[scale=1, nodestyle/.style={draw,circle},baseline=(s0)]

    \node [nodestyle] (s0) at (0,0) {$s_0$};
    \node [] (leftdummy)  [on grid, left=1.2cm of s0] {};
    \node [] (rightdummy) [on grid, right=1.2cm of s0] {};
    \node [nodestyle] (s1) [on grid, below=\distforsone of leftdummy] {$s_1$};
    \node [nodestyle] (s2) [on grid, below=\distforsone of rightdummy] {$s_2$};
    \node [nodestyle, accepting] (s3) [on grid, below=\distsonesthree of s1] {$s_3$};
    \node [nodestyle] (s4) [on grid, below=\distsonesthree of s2] {$s_4$};
    
    \draw ($(s0)-(0.7,0)$) edge[->] (s0);
    \draw (s0) edge[->] node[right] {\scalebox{0.8}{$x^{s_0}$}} (s1);
    \draw (s0) edge[->] node[right] {\scalebox{0.8}{$1{-}x^{s_0}$}} (s2);

     \draw (s1) edge[bend left=15, ->] node[above] {\scalebox{0.8}{$y^{s_1}$}} (s2);
     \draw (s1) edge[->] node[right] {\scalebox{0.8}{$1{-}y^{s_1}$}} (s3);
    
     \draw (s2) edge[bend left=15, ->] node[below] {\scalebox{0.8}{$y^{s_2}$}} (s1);
     \draw (s2) edge[->] node[left] {\scalebox{0.8}{$1{-}y^{s_2}$}} (s4);
    
    \draw (s3) edge[loop right, ->] node[auto] {\scriptsize$1$} (s3);
    
    \draw (s4) edge[loop left, ->] node[auto] {\scriptsize$1$} (s3);
\end{tikzpicture}
    }
    \label{fig:pmc:Dprime}
  }
  \vspace{-2mm}
  \caption{A pMC $\pdtmc$, some instantiation $\pdtmc[u]$ and the relaxation $\rel{\pdtmc}$.}
  \label{fig:pmc}
  \vspace{-4mm}
\end{figure}
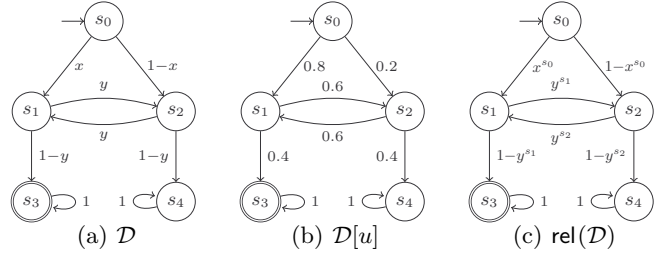

\begin{definition}{\bf (Well-defined region)}\label{def:well_def_region}
Let $\psgGeneric$ be a parametric model.
A region $r$ for $V$ is \emph{well-defined} for $\psgGeneric$ if for all $u\in r$ it holds that
 $u$ is well-defined for $\psgGeneric$, and for all polynomials $f$ in $\psgGeneric$ either $f=0$ or $f[u]>0$.
\end{definition}
The first condition says that $\psgGeneric[u]$ is a probabilistic model (SG, MC, or MDP) while the second one ensures that $\psgGeneric[u]$ and $\psgGeneric$ have the same topology.
\begin{example}
\label{ex:regions}
Let $\pdtmc$ be the pMC in Fig.~\ref{fig:pmc:D}, the region $r=[0.1, 0.8] \times [0.4, 0.7]$ and the valuation $u=(0.8, 0.6) \in r$.
Fig.~\ref{fig:pmc:Du} depicts the instantiation $\pdtmc[u]$, which is an MC as defined in Section~\ref{sec:preliminaries:models} with the same topology as $\pdtmc$.
As this holds for all possible instantiations $\pdtmc[u']$ with $u' \in r$, region $r$ is well-defined.
%
%
The region $r'=[0,1] \times [0,1]$ is not well-defined as, \eg, the valuation $(0,0) \in r'$ results in an MC that has no transition from $s_1$ to $s_2$.
\end{example}
Our aim is to prove that a property $\varphi$ holds \emph{for all instantiations} of a parametric model $\psgGeneric$ which are represented by a region $r$, \ie, $\psgGeneric, r \models \varphi$ defined as follows.
\begin{definition}{\bf (Satisfaction relation for regions)}\label{def:satRel}
For a parametric model $\psgGeneric$, a well-defined region $r$, and a property $\varphi$, the relation $\models$ is defined as
\begin{align*}
\psgGeneric, r \models \varphi \iff \psgGeneric[u] \models \varphi \text{ for all } u\in r.
\end{align*}
\end{definition}
Notice that $\psgGeneric, r \not\models \varphi$ implies  $\psgGeneric[u] \not\models \varphi$ for \emph{some} $u \in r$.
This differs from $\psgGeneric, r \models \neg\varphi$ which implies $\psgGeneric[u] \not\models \varphi$ for \emph{all} $u\in r$.
If $\psgGeneric$ and $\varphi$ are clear from the context, we will call region $r$ \emph{safe} if $\psgGeneric, r \models \varphi$ and \emph{unsafe} if $\psgGeneric, r \models \neg\varphi$.

Let $\pdtmcInit$ be a pMC, $r$ a region that is \emph{well-defined for $\pdtmc$}, and $\reachPropSymbol = \reachProplT$ a reachability property. 
We want to infer that $r$ is safe (or unsafe). We do this by considering the \emph{maximal (or minimal)} possible reachability probability over all valuations $u$ from $r$. 
We give the equivalences for safe regions:
\begin{align*}
\pdtmc, r \models \reachPropSymbol &\iff \big(\max_{u\in r} \reachPr{\pdtmc[u]}{T} \big) \le \lambda\\
\pdtmc, r \models \neg \reachPropSymbol &\iff \big(\min_{u\in r} \reachPr{\pdtmc[u]}{T} \big) > \lambda
\end{align*}
\begin{remark}
As shown in \cite{Daw04}, $\reachPr{\pdtmc[u]}{T}$ can be expressed as a rational function $f = \nicefrac{g_1}{g_2}$ with polynomials $g_1, g_2$.
As $r$ is well-defined, $g_2(u) \neq 0$ for all $u \in r$.
Therefore, $f$ is continuous on the closed set $r$.
Hence, there is always a valuation that induces the maximal (or minimal) reachability probability: 
\begin{align*}
&\sup_{u\in r}\reachPr{\pdtmc[u]}{T}  = 
\max_{u\in r} \reachPr{\pdtmc[u]}{T}\\
 \text{and }& \inf_{u\in r} \reachPr{\pdtmc[u]}{T}  = \min_{u\in r}\reachPr{\pdtmc[u]}{T}\ .	
\end{align*}
\end{remark}
%
\begin{example}
Reconsider the pMC $\pdtmc$ in Fig.~\ref{fig:pmc:D} and region $r = [0.1, 0.8] \times [0.4, 0.7]$. 
We look for a valuation $u \in r$ that maximises  $\pr^{\pdtmc[u]}(\finally \{s_3\})$, \ie, the probability to reach $s_3$ from  $s_0$.
Notice that $s_4$ is the only state from which we cannot reach $s_3$, furthermore, $s_4$ is only reachable via $s_2$.
Hence, it is best to avoid $s_2$.
For the parameter $x$ it follows that the value $u(x)$ should be as high as possible, \ie, $u(x) = 0.8$.
Consider state $s_1$: As we want to reach $s_3$, the value of $y$ should be preferably low.
On the other hand, from $s_2$, $y$ should be assigned a high value as we want to avoid $s_4$.
Thus, it requires a thorough analysis to find an optimal value for $y$, due to the trade-off for the reachability probabilities from $s_1$ and $s_2$.
\end{example}
  
\subsection{Relaxation}
\label{sec:regionsMCs:pmcs}
The idea of our approach, inspired by \cite{BCDS13}, is to drop these dependencies by means of a \emph{relaxation} of the problem in order to ease finding an optimal valuation.
\begin{definition}{\bf (Relaxation)}\label{def:relaxation_pmc}
The \emph{relaxation} of pMC $\pdtmcInit$ is the pMC $\rel{\pdtmc} = (S, \rel[\pdtmc]{V}, \sinit, \probdtmc')$ with
$\rel[\pdtmc]{V}=\{x_i^s \mid x_i \in \Var, s\in S\}$ and  $\probdtmc'(s,s')=\probdtmc(s,s')[x_1, \dots, x_n / x_1^s, \dots, x_n^s]$.
\end{definition}
Intuitively, the relaxation $\rel{\pdtmc}$ arises from $\pdtmc$ by equipping each state with its own parameters and thereby eliminating parameter dependencies.
We extend a valuation $u$ for $\pdtmc$ to the \emph{relaxed valuation} $\rel[\pdtmc]{u}$ for $\rel{\pdtmc}$ by $\rel[\pdtmc]{u}(x_i^s) = u(x_i)$ for every $s$. We have that for all $u$, $\pdtmc[u] = \pdtmc[\rel[\pdtmc]{u}]$.
We lift the relaxation to regions such that $B(x_i^s) = B(x_i)$ for all $s$, \ie, $\rel[\pdtmc]{r} = \bigtimes_{x_i^s \in \rel[\pdtmc]{V}} B(x_i)$.
We drop the subscript $\pdtmc$, whenever it is clear from the context.
\begin{example}\label{ex:relaxation}
Fig.~\ref{fig:pmc:Dprime} depicts the relaxation $\rel{\pdtmc}$ of the pMC $\pdtmc$ from Fig.~\ref{fig:pmc:D}.
For $r=[0.1, 0.8] \times [0.4, 0.7]$ and $u=(0.8, 0.6) \in r$ from Example~\ref{ex:regions}, we obtain $\rel{r}=[0.1, 0.8] \times [0.4, 0.7] \times [0.4, 0.7]$ and $\rel{u}=(0.8, 0.6, 0.6)$.
The instantiation $\rel{\pdtmc}[\rel{u}]$ corresponds to $\pdtmc[u]$ as depicted in Fig.~\ref{fig:pmc:Du}.
Notice that the relaxed region $\rel{r}$ contains also valuations, e.g., $(0.8, 0.5, 0.6)$ which give rise to instantiations which are not realisable by valuations in $r$. 
\end{example}
For a pMC $\pdtmc$ and a region $r$ that is well-defined for $\pdtmc$, notice that $\{\pdtmc[u] \mid u \in r\} \subseteq \{\rel{\pdtmc}[u] \mid u \in \rel{r}\}$. 
Due to the fact that $\rel{\pdtmc}$ is an over-approximation of $\pdtmc$, the maximal reachability probability over all instantiations of $\pdtmc$ within $r$ is at most as high as the one for all instantiations of $\rel{\pdtmc}$ within $\rel{r}$. 
\begin{lemma}\label{lem:relaxationOverapproximates}
For pMC $\pdtmc$ and well-defined region $r$, we have
\begin{align*}
 \max_{u\in r}\big( \reachPr{\pdtmc[u]}{T} \big) \ = \
 \max_{u\in r}\big( \reachPr{\rel{\pdtmc}[\rel{u}]}{T} \big) \ \le \
 \max_{u\in \rel{r}}\big( \reachPr{\rel{\pdtmc}[u]}{T} \big).
\end{align*}
\end{lemma}
%
Thus, if the relaxation satisfies a reachability property, so does the original pMC.
\begin{corollary}
Given a pMC $\pdtmc$ and a well-defined region $r$ it holds that
\begin{align*}
\max_{u\in \rel{r}}\big( \reachPr{\rel{\pdtmc}[u]}{T}\big) \le \lambda \text{ implies }\pdtmc, r \models \reachPropSymbol.
\end{align*}
\end{corollary}
Note that the relaxation does not aggravate the problem for our setting. 
In fact, although \rel{\pdtmc} has (usually) much more parameters than $\pdtmc$, it is intuitively easier to find a valuation $u \in \rel{r}$ that maximises the reachability probability:
For some $x_i^s \in \rel{V}$, we can always pick a value in $I(x^s_i)$ that maximises the probability to reach $T$ from state $s$.
There is no (negative) effect for the reachability probability at the remaining states as $x_i^s$ only occurs at $s$.

Recall that the functions $f$ occurring in $\rel{\pdtmc}$ are of the form $f=\sum_{i \le m} a_i \cdot \prod_{x\in V_i} x$ (with $a_i \in \Q$ and $V_i \subseteq \rel{V}$).
Finding a valuation that maximises the reachability probability becomes especially easy for this setting:
We only need to consider valuations $u$ that set the value of each parameter to either the lowest or highest possible value, \ie, $u(x_i^s) \in B(x_i^s)$ for all $x_i^s \in \rel{V}$.
This important result is stated as follows.
\begin{theorem}
\label{lem:optimalValuation}
Let $\pdtmc$ be a pMC, $r$ be a well-defined region, and $T\subseteq S$ be a set of target states.
 There is a valuation $u' \in \rel{r}$ satisfying $u'(x_i^s) \in B(x_i^s)$ for all $x_i^s\in \rel{V}$ such that $\reachPr{\rel{\pdtmc}[u']}{T} = \max_{u\in \rel{r}}\reachPr{\rel{\pdtmc}[u]}{T}$.
\end{theorem}
We prove this by showing that any valuation which assigns some variable to something other than its bound can be modified such that the variable is assigned to its bound, without decreasing the induced reachability probability. 
\begin{proof}
Let $u'\in \rel{r}$ with $\reachPr{\rel{\pdtmc}[u']}{T} = \max_{u\in \rel{r}}\big( \reachPr{\rel{\pdtmc}[u]}{T} \big)$. For the sake of contradiction,  assume that there exists a parameter $x_i^s \in \rel{\Var}$ with $u'(x_i^s) \in I(x_i^s) \setminus B(x_i^s)$ such that for all $u'' \in r$ with $u''(y) = u'(y)$ for all $y \in \rel{\Var} \setminus \{ x_i^s \}$ and $u''(x_i^s) \in B(x_i^s)$ it holds that
$\reachPr{\rel{\pdtmc}[u']}{T} > \reachPr{\rel{\pdtmc}[u'']}{T}$.
We show that no such $x_i^s$ exists. 
%

W.\,l.\,o.\,g. let $s\notin T$ be the initial state of $\rel{\pdtmc}$ and let $T$ be reachable from $s$.
Consider the pMC $\hat{\pdtmc} = (S, \{x_i^s\}, s, \hat{\probdtmc})$ with the single parameter $x_i^s$ that arises from $\rel{\pdtmc}$ by replacing all parameters $x \in \rel{\Var}\setminus\{x_i^s\}$ with $u'(x)$.
Let $\pctlUntil$ denote the standard until-modality and $\neg T$ denote $S\setminus T$.
Using the characterisation of reachability probabilities as linear equation system (cf.~\cite{BK08}), the reachability probability \wrt $T$ in $\hat{D}$ is given by
\begin{align*}
 \reachPrT[\hat{\pdtmc}] 
=& \sum_{s'\in S} \hat{\probdtmc}(s,s') \cdot \reachPrs{\pdtmc}{s'}{T} \\
=& \sum_{s'\in S} \hat{\probdtmc}(s,s') \cdot \Big(\pr_{s'}^{\hat{\pdtmc}}(\neg s \, \pctlUntil \, T) + \pr_{s'}^{\hat{\pdtmc}}(\neg T \, \pctlUntil \, s) \cdot \reachPrT[\hat{\pdtmc}] \Big)\\
=& \underbrace{\sum_{s'\in S} \overbrace{\hat{\probdtmc}(s,s')}^{\mathit{linear}} \cdot \overbrace{\pr_{s'}^{\hat{\pdtmc}}(\neg s \, \pctlUntil \, T)}^{\mathit{constant}}}_{\eqqcolon f}  +  \underbrace{\sum_{s'\in S}    \overbrace{\hat{\probdtmc}(s,s')}^{\mathit{linear}} \cdot \overbrace{\pr_{s'}^{\hat{\pdtmc}}(\neg T \, \pctlUntil \, s)}^{\mathit{constant}}}_{\eqqcolon g} \cdot \reachPrT[\hat{\pdtmc}]
\end{align*}
Note that $f$ and $g$ are linear functions over the parameter $x_i^s$.
Furthermore, $g(a) < 1$ for all $a\in I(x_i^s)$ as $T$ is reachable from $s$.
Transposing the equation yields
$
\reachPrT[\hat{\pdtmc}] = \nicefrac{f}{(1-g)}. 
$
Note that this function is monotonic on $r$.
Thus, we can set  $u'(x_i^s)$ to a value in $B(x_i^s)$  without decreasing the reachability probability.
This contradicts our assumption that $u'(x_i^s) \in B(x_i^s)$ holds for a maximum number of parameters $x_i^s \in \rel{V}$. \qed\end{proof}

\begin{example}
	Let $\rel{\pdtmc}$ as in Fig.~\ref{fig:pmc:Dprime} and $r=[0.1, 0.8] \times [0.4, 0.7]$. Considering $u' \in \rel{r}$ with $u'(y^{s_1}) = u'(y^{s_2}) = 0.5$ we obtain the linear functions
	\[
        f(x^{s_0}) = \frac{2}{3} x^{s_0} + \frac{1}{3} (1 - x^{s_0}) = \frac{1}{3} x^{s_0} + \frac{1}{3} \qquad \text{ and } \qquad g(x^{s_0}) = 0.
  	\]
  	Note that $\nicefrac{f}{1 - g} = f$ is montonically increasing and the reachability probability can therefore be increased by setting $x^{s_0}$ to its upper bound $0.7$.
\end{example}

  \subsection{Substituting Parameters with Nondeterminism}
\label{sec:regionsMCs:substituting}
We have now seen that, in order to determine $\max_{u\in \rel{r}	} \reachPr{\rel{\pdtmc}[u]}{T}$,
we have to make a discrete choice over valuations $v \colon \rel{\Var} \rightarrow \R$ with $v(x_i^s) \in B(x_i)$. 
This choice can be made locally at every state, which brings us to the key idea of \emph{constructing a (non-parametric) MDP out of the pMC $\pdtmc$ and the region $r$}, where nondeterministic choices represent all valuations that need to be considered.

\begin{figure}[t]
  \centering
  \subfigure[$\pdtmc$]{
    \scalebox{\picscale}{
      \begin{tikzpicture}[scale=1, nodestyle/.style={draw,circle},baseline=(s0)]
    
    \node [nodestyle] (s0) at (0,0) {$s_0$};
    \node [] (leftdummy)  [on grid, left=1.2cm of s0] {};
    \node [] (rightdummy) [on grid, right=1.2cm of s0] {};
    \node [nodestyle] (s1) [on grid, below=\distforsone of leftdummy] {$s_1$};
    \node [nodestyle] (s2) [on grid, below=\distforsone of rightdummy] {$s_2$};
    \node [nodestyle, accepting] (s3) [on grid, below=\distsonesthree of s1] {$s_3$};
    \node [nodestyle] (s4) [on grid, below=\distsonesthree of s2] {$s_4$};
    
    \draw ($(s0)-(0.7,0)$) edge[->] (s0);
    \draw (s0) edge[->] node[right] {\scriptsize$x$} (s1);
    \draw (s0) edge[->] node[right] {\scriptsize$1{-}x$} (s2);

     \draw (s1) edge[bend left=15, ->] node[above] {\scriptsize$y$} (s2);
     \draw (s1) edge[->] node[right] {\scriptsize$1{-}y$} (s3);
    
     \draw (s2) edge[bend left=15, ->] node[below] {\scriptsize$y$} (s1);
     \draw (s2) edge[->] node[left] {\scriptsize$1{-}y$} (s4);
    
    \draw (s3) edge[loop right, ->] node[auto] {\scriptsize$1$} (s3);
    
    \draw (s4) edge[loop left, ->] node[auto] {\scriptsize$1$} (s3);

\end{tikzpicture}
    }
    \label{fig:approxmdp:D}
  }
  \subfigure[$\subst{\pdtmc}{r}$]{
    \scalebox{\picscale}{
      \begin{tikzpicture}[scale=1, nodestyle/.style={draw,circle},baseline=(s0)]
   
    \node [nodestyle] (s0) at (0,0) {$s_0$};
    \node [] (leftdummy)  [on grid, left=1.2cm of s0] {};
    \node [] (rightdummy) [on grid, right=1.2cm of s0] {};
    \node [nodestyle] (s1) [on grid, below=\distforsone of leftdummy] {$s_1$};
    \node [nodestyle] (s2) [on grid, below=\distforsone of rightdummy] {$s_2$};
    \node [nodestyle, accepting] (s3) [on grid, below=\distsonesthree  of s1] {$s_3$};
    \node [nodestyle] (s4) [on grid, below=\distsonesthree of s2] {$s_4$};
    
    \draw ($(s0)-(0.7,0)$) edge[->] (s0);
    \draw (s0) edge[->, bend left=0, dashed] node[left] {\scriptsize$0.1$} (s1);
    \draw (s0) edge[->, bend right=0, dashed] node[right] {\scriptsize$0.9$} (s2);
    \draw (s0) edge[->, bend right=40] node[left] {\scriptsize$0.8$} (s1);
    \draw (s0) edge[->, bend left=40] node[right] {\scriptsize$0.2$} (s2);

     \draw (s1) edge[bend left=40, ->, dashed] node[above] {\scriptsize$0.4$} (s2);
     \draw (s1) edge[->, bend left=15, dashed] node[right] {\scriptsize$0.6$} (s3);
     \draw (s1) edge[bend left=10, ->] node[above] {\scriptsize$0.7$} (s2);
     \draw (s1) edge[->, bend right=15] node[left] {\scriptsize$0.3$} (s3);
    
     \draw (s2) edge[bend left=40, ->, dashed] node[below] {\scriptsize$0.4$} (s1);
     \draw (s2) edge[->, bend right=15, dashed] node[left] {\scriptsize$0.6$} (s4);
     \draw (s2) edge[bend left=10, ->] node[below] {\scriptsize$0.7$} (s1);
     \draw (s2) edge[->, bend left=15] node[right] {\scriptsize$0.3$} (s4);
    
    \draw (s3) edge[loop right, ->] node[auto] {\scriptsize$1$} (s3);
    
    \draw (s4) edge[loop left, ->] node[auto] {\scriptsize$1$} (s3);
\end{tikzpicture}
    }
    \label{fig:approxmdp:M}
  }
  \subfigure[$\subst{\pdtmc}{r}^\sched$]{
    \scalebox{\picscale}{
      \begin{tikzpicture}[scale=1, nodestyle/.style={draw,circle},baseline=(s0)]
   
    \node [nodestyle] (s0) at (0,0) {$s_0$};
    \node [] (leftdummy)  [on grid, left=1.2cm of s0] {};
    \node [] (rightdummy) [on grid, right=1.2cm of s0] {};
    \node [nodestyle] (s1) [on grid, below=\distforsone of leftdummy] {$s_1$};
    \node [nodestyle] (s2) [on grid, below=\distforsone of rightdummy] {$s_2$};
    \node [nodestyle, accepting] (s3) [on grid, below=\distsonesthree of s1] {$s_3$};
    \node [nodestyle] (s4) [on grid, below=\distsonesthree of s2] {$s_4$};
    
    \draw ($(s0)-(0.7,0)$) edge[->] (s0);
    \draw (s0) edge[->,] node[left] {\scriptsize$0.8$} (s1);
    \draw (s0) edge[->, ] node[right] {\scriptsize$0.2$} (s2);

     \draw (s1) edge[bend left=15, ->, dashed] node[above] {\scriptsize$0.4$} (s2);
     \draw (s1) edge[->, dashed] node[right] {\scriptsize$0.6$} (s3);
    
     \draw (s2) edge[bend left=15, ->] node[below] {\scriptsize$0.7$} (s1);
     \draw (s2) edge[->] node[right] {\scriptsize$0.3$} (s4);
    
    \draw (s3) edge[loop right, ->] node[auto] {\scriptsize$1$} (s3);
    
    \draw (s4) edge[loop left, ->] node[auto] {\scriptsize$1$} (s3);
\end{tikzpicture}
    }
    \label{fig:approxmdp:Msched}
  }
    \vspace{-2mm}
  \caption{Illustrating parameter-substitution.}
  \label{fig:approxmdp}
  \vspace{-4mm}
\end{figure}
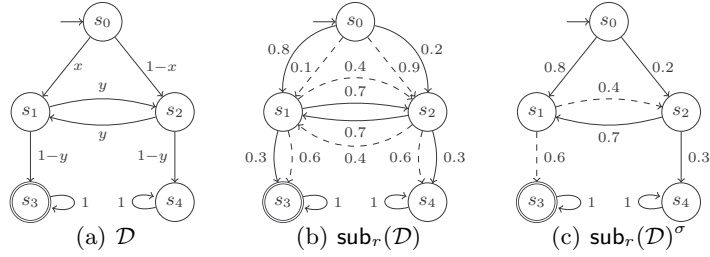

\begin{definition}{\bf (Substitution{-}pMC)}
	\label{def:approxPMC}
An MDP $\subst{\pdtmc}{r} = (S, \sinit, \Act_{\substAbbrev}, \probmdp_{\substAbbrev})$ is the \emph{(parameter-)substitution of a pMC $\pdtmcInit$ and a region $r$} if $\Act_{\substAbbrev} = \biguplus_{s \in S}  \{v \colon \Var_s \rightarrow \R \mid v(x_i) \in B(x_i)\}$
and
\[
\probmdp_{\substAbbrev}(s,v, s') = 
\begin{cases}
\probdtmc(s,s')[v] &\text{if } v \in \Act_s,\\
0                    &\text{otherwise.}
\end{cases}
\]
\end{definition}
Thus, choosing action $v$ in $s$ corresponds to assigning the extremal values $B(x_i)$ to the parameters $x_i^s$. The number of outgoing actions for $s$ is therefore $2^{|\Var_s|}$. 
\begin{example}
\label{ex:approxmdp}
Consider pMC $\pdtmc$ -- depicted in Fig.~\ref{fig:approxmdp:D}
-- with $r = [0.1, 0.8] \times [0.4, 0.7]$ as before. The substitution of $\pdtmc$ on $r$ is shown in Fig.~\ref{fig:approxmdp:M}.
In $\pdtmc$, each outgoing transition of states $s_0, s_1, s_2$ is replaced by a nondeterministic choice in $\subst{\pdtmc}{r}$. That is, we either pick the upper or lower bound for the corresponding variable.
The solid (dashed) lines depict transitions that belong to the action for the upper (lower) bound.
For the states $s_3$ and $s_4$ there is no choice, as their outgoing transitions in $\pdtmc$ are constant.
Fig.~\ref{fig:approxmdp:Msched} depicts the MC $\subst{\pdtmc}{r}^\sched$ which is induced by the scheduler $\sched$ on $\subst{r}{\pdtmc}$ that chooses the upper bounds at $s_0$ and $s_2$, and the lower bound at $s_1$.
Notice that $\subst{\pdtmc}{r}^\sched$ coincides with $\rel{\pdtmc}[v]$ for a suitable valuation $v$, as depicted in Fig. \ref{fig:pmc:Dprime}.
\end{example}
%
First, observe that the nondeterministic choices introduced by the substitution only depend on the values $B(x_i)$ of the parameters $x_i$ in $r$. Since the ranges of the parameters $x_i^s$ in $\rel{r}$ agree with the range of $x_i$ in $r$, we have 
\begin{align}
	\subst{\rel{\pdtmc}}{\rel{r}} = \subst{\pdtmc}{r} \quad \text{for all well-defined } r.
	\label{eqn:relaxationNotNecessary}
\end{align} 
%
Second, note that the substitution encodes the local choices for a relaxed pMC. That is, for an arbitrary pMC, there is a one-to-one correspondence between schedulers $\sched \in \Sched{\subst{\rel{\pdtmc}}{\rel{r}}}$ and valuations $v \in \rel{r}$ for $\rel{\pdtmc}$ with $v(x_i^s) \in B(x_i)$.
Combining the observations with Theorem~\ref{lem:optimalValuation}, yields the following. 
 \begin{corollary}\label{cor:dtmcApprox}
For a pMC $\pdtmc$, a region $r$ and a set of target states $T$ of $\pdtmc$:
\begin{align*}
 \max_{u\in r} \reachPrT[{\pdtmc[u]}] \; \le \; 
\smashoperator{\max_{\sched \in \SchedSymbol}} \reachPrT[\subst{\rel{\pdtmc}}{\rel{r}}^\sched] \; = \;
\smashoperator{\max_{\sched \in \SchedSymbol}} \reachPrT[\subst{\pdtmc}{r}^\sched]  \\
 \min_{u\in r} \reachPrT[{\pdtmc[u]}] \; \ge \;
\smashoperator{\min_{\sched \in \SchedSymbol}} \reachPrT[\subst{\rel{\pdtmc}}{\rel{r}}^\sched] \; = \;
\smashoperator{\min_{\sched \in \SchedSymbol}} \reachPrT[\subst{\pdtmc}{r}^\sched] 
\end{align*}
\end{corollary}
As a direct consequence of this, we can state Theorem~\ref{lem:dtmcApprox}.
\begin{theorem}
\label{lem:dtmcApprox}
Let $\pdtmc$ be a pMC, $r$ be a well-defined region. Then
\begin{align*}
 \subst{\pdtmc}{r} \models \reachProplT \text{ implies }& \pdtmc, r \models \reachProplT  \text{ and }\\
 \subst{\pdtmc}{r} \models \p_{> \lambda}(\finally T)\text{ implies }& \pdtmc, r \models \neg \reachProplT .
\end{align*}
\end{theorem}
Hence, we can deduce whether $\pdtmc, r \models \varphi$ by applying standard techniques for MDP model checking to $\subst{\pdtmc}{r}$. If the over-approximation is too coarse for a conclusive answer, regions can be refined (cf.~Section~\ref{sec:synthesis}).
Moreover, while the relaxation is key for showing the correctness, equation (\ref{eqn:relaxationNotNecessary}) proves that this step does not actually need to be performed.
\begin{example}
  Reconsider Example \ref{ex:approxmdp}. From $\subst{\pdtmc}{r}$ in Fig. \ref{fig:approxmdp:M}, we can derive $\max_{\sched \in \SchedSymbol} \reachPrT[\subst{\pdtmc}{r}^\sched] = \nicefrac{47}{60}$ and, by Theorem \ref{lem:dtmcApprox}, $\pdtmc, r \models \reachProp{0.8}{T}$ follows. Despite the large considered region, we were able to establish a non-trivial upper bound on the reachability probability over all valuations in $r$.
\end{example}
\medskip\noindent\emph{Expected Reward Properties.}
The notions above can be applied to perform regional model checking of pMCs and expected reward properties.
Regions have to be further restricted such that: $\reachPr{\pdtmc[u]}{T} = 1$ for all $u \in r$  -- to ensure that the expected reward is defined -- and, for transition-rewards, reward-parameters and probability-parameters have to be disjoint.
We can then generalise relaxation and substitution to the reward models, and obtain analogous results.

  \section{Regional Checking of Models with Nondeterminism}
\label{sec:nondet}
In the last section we showed how to bound reachability probabilities of pMCs from below and above. Introducing nondeterministic choices between these bounds enabled to utilise standard MDP model checking for the parameter synthesis.
This approach can readily be generalised to systems originally exhibiting nondeterminism. In particular, for pMDPs this adds choices over valuations (inherent to parameters) to the choices over actions (inherent to MDPs).
This new nondeterminism leads to a game with two players: One for the nondeterminism of the MDP and one for the abstracted parameters, yielding a stochastic game.

In the following, let $\pmdpInit$ be a pMDP and $r$ a well-defined region for $\pmdp$.
We want to analyse $r$ for all scheduler-induced pMCs $\pmdp^\sched$ of $\pmdp$. 
\begin{figure}[t]
	\centering
	\subfigure[$\pmdp$]{
		\scalebox{0.85}{
			\begin{tikzpicture}[scale=1, nodestyle/.style={draw,circle, inner sep=1.5pt},baseline=(s)]
    
    \node [nodestyle] (s) at (0,0) {$s$};
    \node [draw, fill, circle, inner sep=0.5pt] (sa) [on grid, above right=1cm of s] {};
    \node [draw, fill, circle, inner sep=0.5pt] (sb) [on grid, below right=1cm of s] {};
    \node [nodestyle] (sa1) at ($(sa) + (1,0.5)$) {};
    \node [nodestyle] (sa2) at ($(sa) + (1,-0.5)$) {};
    \node [nodestyle] (sb1) at ($(sb) + (1,0.5)$) {};
    \node [nodestyle] (sb2) at ($(sb) + (1,-0.5)$) {};
    
    \draw (s) edge[] node[above, pos=0.25] {\scriptsize$\alpha$} (sa);
    \draw (s) edge[] node[below, pos=0.25] {\scriptsize$\beta$} (sb);
    
    \draw (sa) edge[->] node[right] {\scriptsize} (sa1);
    \draw (sa) edge[->] node[right] {\scriptsize} (sa2);
    \draw (sb) edge[->] node[right] {\scriptsize} (sb1);
    \draw (sb) edge[->] node[right] {\scriptsize} (sb2);
    
\end{tikzpicture}
		}
		\label{fig:pmdp:M}
	}
	\subfigure[$\subst{\pmdp^\sched}{r}$]{
		\scalebox{0.85}{
			\begin{tikzpicture}[scale=1, nodestyle/.style={draw,circle, inner sep=1.5pt},baseline=(s)]

\node [nodestyle] (s) at (0,0) {$s$};
\node [nodestyle] (sa1) at ($(sa) + (1,0.5)$) {};
\node [nodestyle] (sa2) at ($(sa) + (1,-0.5)$) {};
\node [nodestyle] (sb1) at ($(sb) + (1,0.5)$) {};
\node [nodestyle] (sb2) at ($(sb) + (1,-0.5)$) {};

\draw (s) edge[->, bend left=25] node[right] {\scriptsize} (sa1);
\draw (s) edge[->, bend right=25] node[right] {\scriptsize} (sa2);
\draw (s) edge[->, dashed, bend right=0] node[right] {\scriptsize} (sa1);
\draw (s) edge[->, dashed, bend left=0] node[right] {\scriptsize} (sa2);

\end{tikzpicture}
		}
		\label{fig:pmc:Msched}
	}
	\subfigure[$\pmdp'$]{
		\scalebox{0.85}{
			\begin{tikzpicture}[scale=1, nodestyle/.style={draw,circle, inner sep=1.5pt},baseline=(s)]
    
    \node [nodestyle] (s) at (0,0) {$s$};
    \node [nodestyle, rectangle, minimum height=12pt] (sa) [on grid, above right=1cm of s] {$s,\alpha$};
    \node [nodestyle, rectangle, minimum height=12pt] (sb) [on grid, below right=1cm of s] {$s,\beta$};
    \node [nodestyle] (sa1) at ($(sa) + (1,0.5)$) {};
    \node [nodestyle] (sa2) at ($(sa) + (1,-0.5)$) {};
    \node [nodestyle] (sb1) at ($(sb) + (1,0.5)$) {};
    \node [nodestyle] (sb2) at ($(sb) + (1,-0.5)$) {};
    
    \draw (s) edge[->] node[above, pos=0.25] {\scriptsize$\alpha$} (sa);
    \draw (s) edge[->] node[below, pos=0.25] {\scriptsize$\beta$} (sb);
    
    \draw (sa) edge[->] node[right] {\scriptsize} (sa1);
    \draw (sa) edge[->] node[right] {\scriptsize} (sa2);
    \draw (sb) edge[->] node[right] {\scriptsize} (sb1);
    \draw (sb) edge[->] node[right] {\scriptsize} (sb2);

\end{tikzpicture}
		}
		\label{fig:pmdp:Mp}
	}
	\subfigure[$\psg$]{
		\scalebox{0.85}{
			\begin{tikzpicture}[scale=1, nodestyle/.style={draw,circle, inner sep=1.5pt},baseline=(s)]
    
    \node [nodestyle] (s) at (0,0) {$s$};
    \node [nodestyle, rectangle, minimum height=12pt] (sa) [on grid, above right=1cm of s] {$s,\alpha$};
    \node [nodestyle, rectangle, minimum height=12pt] (sb) [on grid, below right=1cm of s] {$s,\beta$};
    \node [nodestyle] (sa1) at ($(sa) + (1,0.5)$) {};
    \node [nodestyle] (sa2) at ($(sa) + (1,-0.5)$) {};
    \node [nodestyle] (sb1) at ($(sb) + (1,0.5)$) {};
    \node [nodestyle] (sb2) at ($(sb) + (1,-0.5)$) {};
    
    \draw (s) edge[->] node[above, pos=0.25] {\scriptsize$\alpha$} (sa);
    \draw (s) edge[->] node[below, pos=0.25] {\scriptsize$\beta$} (sb);
    
    \draw (sa) edge[->, bend left=15] node[right] {\scriptsize} (sa1);
    \draw (sa) edge[->, bend right=15] node[right] {\scriptsize} (sa2);
    \draw (sb) edge[->, bend left=15] node[right] {\scriptsize} (sb1);
    \draw (sb) edge[->, bend right=15] node[right] {\scriptsize} (sb2);
    
    \draw (sa) edge[->, dashed, bend right=15] node[right] {\scriptsize} (sa1);
    \draw (sa) edge[->, dashed, bend left=15] node[right] {\scriptsize} (sa2);
    \draw (sb) edge[->, dashed, bend right=15] node[right] {\scriptsize} (sb1);
    \draw (sb) edge[->, dashed, bend left=15] node[right] {\scriptsize} (sb2);
    
\end{tikzpicture}
		}
		\label{fig:pmdp:S}
	}
	\subfigure[$\psg^\sched$]{
		\scalebox{0.85}{
			\begin{tikzpicture}[scale=1, nodestyle/.style={draw,circle, inner sep=1.5pt},baseline=(s)]

\node [nodestyle] (s) at (0,0) {$s$};
\node [nodestyle, rectangle, minimum height=12pt] (sa) [on grid, above right=1cm of s] {$s,\alpha$};
\node [nodestyle, rectangle, minimum height=12pt] (sb) [on grid, below right=1cm of s] {$s,\beta$};
\node [nodestyle] (sa1) at ($(sa) + (1,0.5)$) {};
\node [nodestyle] (sa2) at ($(sa) + (1,-0.5)$) {};
\node [nodestyle] (sb1) at ($(sb) + (1,0.5)$) {};
\node [nodestyle] (sb2) at ($(sb) + (1,-0.5)$) {};

\draw (s) edge[->] node[above, pos=0.25] {\scriptsize$\alpha$} (sa);

\draw (sa) edge[->, bend left=15] node[right] {\scriptsize} (sa1);
\draw (sa) edge[->, bend right=15] node[right] {\scriptsize} (sa2);
\draw (sb) edge[->, bend left=15] node[right] {\scriptsize} (sb1);
\draw (sb) edge[->, bend right=15] node[right] {\scriptsize} (sb2);

\draw (sa) edge[->, dashed, bend right=15] node[right] {\scriptsize} (sa1);
\draw (sa) edge[->, dashed, bend left=15] node[right] {\scriptsize} (sa2);
\draw (sb) edge[->, dashed, bend right=15] node[right] {\scriptsize} (sb1);
\draw (sb) edge[->, dashed, bend left=15] node[right] {\scriptsize} (sb2);

\end{tikzpicture}
		}
		\label{fig:pmdp:Ssched}
	}
	\vspace{-2mm}
	\caption{Illustration of the substitution of a pMDP.}
	\label{fig:pmdp}
		\vspace{-4mm}
\end{figure}
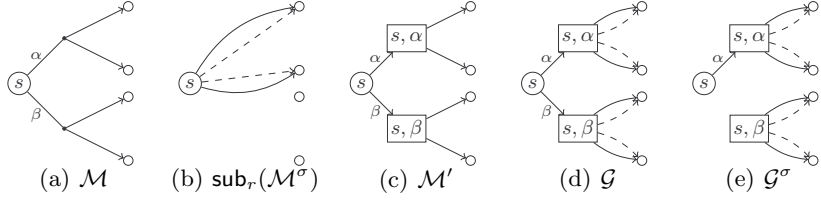

\begin{example}\label{ex:substmdp}
Consider the pMDP $\pmdp$ in Fig.~\ref{fig:pmdp:M}, where state $s$ has two enabled actions $\alpha$ and $\beta$.
The scheduler $\sigma$ given by $\{s \mapsto \alpha\}$ applied to $\pmdp$ yields a pMC, which is subject to substitution, cf.\ Fig.~\ref{fig:pmc:Msched}.
\end{example}
The parameter substitution of a pMDP (cf.~Fig.~\ref{fig:pmdp:M}) yields an SG---as in Fig.~\ref{fig:pmdp:S}. It represents, for all schedulers of the pMDP, the substitution of each induced pMC. For the construction of the substitution, we first introduce intermediate states to separate nondeterministic actions from probabilistic choices in two steps:
\begin{compactitem}
\item Split each state $s \in S$ into $\{s\} \uplus \{\langle s, \alpha \rangle \mid \alpha \in \Act(s) \}$.
\item For $s \in S$ and $\alpha \in \Act(s)$, add a transition with probability one from $s$ to $\langle s, \alpha \rangle$ and move the probabilistic choice at $s$ \wrt $\alpha$ to $\langle s, \alpha \rangle$.  	
\end{compactitem}
We obtain a pMDP as in Fig.~\ref{fig:pmdp:Mp} where state $s$ has pure \emph{nondeterministic} choices leading to states of the form $\langle s, \alpha \rangle$ with pure \emph{probabilistic} choices.
The subsequent substitution on the probabilistic states yields the  stochastic game, where one player represents the nondeterminism of the original pMDP, while the other player decides whether parameters should be set to their lower or upper bound. 
Formally, the game $\sg = \subst{\pmdp}{r}$ is defined as follows.
\begin{definition}{\bf (Substitution{-}pMDP)}
	\label{def:approxPMDP} 
	Given a pMDP $\pmdpInit$ and a region $r$, 
	an SG $\subst{\pmdp}{r} = (S_{\pOne} \uplus S_{\pTwo}, \sinit, \Act_{\substAbbrev}, \probsg_{\substAbbrev})$ with $S_{\pOne} = S$ and $S_{\pTwo} = \{\langle s, \alpha \rangle \mid \alpha \in \Act(s) \}$
	 is the \emph{(parameter-)substitution of $\pmdp$ and $r$}  if	$\Act_{\substAbbrev} = \Act \uplus \big(\biguplus_{\langle s, \alpha \rangle \in \spTwo} \Act_s^\alpha\big)$ with $\Act_s^\alpha = \{ v \colon V_{s}^\alpha \rightarrow \R \mid v(x_i) \in B(x_i) \}$ and
	\[
	\probsg_{\substAbbrev}(t,\beta,t') = 
	\begin{cases}
	1                                        & \text{if } t \in \spOne \text{ and } t' = \langle t,\beta\rangle \in \spTwo,\\
	\probmdp(s,\alpha,t')[\beta]                & \text{if } t = \langle s,\alpha \rangle \in \spTwo, \beta \in \Act_s^\alpha, \text{and } t' \in \spOne, \\
	0                                        & \text{otherwise.}
	\end{cases}
	\]
\end{definition}
%
We now relate the obtained stochastic game $\sg = \subst{\pmdp}{r}$ under different schedulers for player $\pOne$ with the substitution in the scheduler-induced pMCs of $\pmdp$. We observe that the schedulers $\sched \in \Sched[\pOne]{\sg}$ for player $\pOne$ coincide with the schedulers in $\pmdp$.
Consider $\sg^\sched$ with $\sched \in \Sched[\pOne]{\sg}$ which arises from $\sg$ by erasing transitions not agreeing with $\sigma$, \ie, we set all $\probsg_{\sg}(s, \alpha, \langle s, \alpha \rangle)$ with $s \in \spOne$ and $\alpha \neq \sched(s)$ to zero.
Note that $\sg^\sched$ is an MDP as at each state of player $\pOne$, only one action is enabled and therefore only player $\pTwo$ has nondeterministic choices.
\begin{example}
Continuing Example~\ref{ex:substmdp}, applying scheduler $\sigma$ to $\sg$ yields $\sg^\sigma$, see Fig.~\ref{fig:pmdp:Ssched}. The MDP $\sg^\sched$ matches the MDP $\subst{\pmdp^{\sched}}{r}$
apart from intermediate states of the form $\langle s, \alpha \rangle$:
 The state $s$ in $\subst{\pmdp^\sigma}{r}$ has the same outgoing transitions as the state $\langle s, \alpha \rangle$ in $\sg^\sched$ and $\langle s, \alpha \rangle$ is the unique successor of $s$ in $\sg^\sched$.
\end{example}
Note that $\sg^\sched$ and $\subst{\pmdp^{\sched}}{r}$ induce the same reachability probabilities. Formally:
\begin{corollary}
	\label{cor:mdpApprox}
	For pMDP $\pmdp$, well-defined region $r$, target states $T\in S$, and schedulers $\sched \in \Sched[\pOne]{\subst{\pmdp}{r}}$ and $\altsched \in \Sched{\subst{\pmdp^\sigma}{r}}$, it holds that
	\begin{align*}
		\reachPrT[(\subst{\pmdp^\sigma}{r})^\altsched] 
	=
	\reachPrT[\subst{\pmdp}{r}^{\sched, \widehat{\altsched}}]	
	\end{align*}
	with $\widehat{\altsched} \in \Sched[\pTwo]{\subst{\pmdp}{r}}$ 
	satisfies $\widehat{\altsched}(\langle s, \sched(s) \rangle) = \altsched(s)$.
\end{corollary}
Instead of performing the substitution on the pMC induced by $\pmdp$ and $\sched$, we can perform the substitution on $\pmdp$ directly and preserve the reachability probability.
\begin{theorem}
\label{th:mdpApprox}
Let $\pmdp$ be a pMDP, $r$ be a well-defined region. Then
\begin{align*}
\subst{\pmdp}{r} \models_\emptyset \reachProplT \text{ implies }& \pmdp, r \models \reachProplT  \text{, and }\\
\subst{\pmdp}{r} \models_{\{\pOne\}} \p_{> \lambda}(\finally T)\text{ implies }& \pmdp, r \models \neg \reachProplT .	
\end{align*}
\end{theorem}
Therefore, analogously to the pMC case (cf. Theorem~\ref{lem:dtmcApprox}), we can derive whether $\subst{\pmdp}{r} \models \varphi$ by analysing a stochastic game. The formal proof is in the appendix.
%


  \section{Parameter Synthesis}
\label{sec:synthesis}
In this section we briefly discuss how the regional model checking is embedded into a complete parameter space partitioning framework as, \eg, described in~\cite{dehnert-et-al-cav-2015}.
The goal is to partition the parameter space into \emph{safe} and \emph{unsafe} regions (cf.~ Section~\ref{sec:regionsMC:regions}). From a practical point of view, yielding a 100\% coverage of the parameter space is not realistic; 
instead a large coverage (say, 95\%) is aimed at.

We discuss the complete chain for a pMDP $\mdp$ and a property $\varphi$. In addition, a well-defined region $R$ is given which serves as \emph{parameter space}. Recall that a region $r\subseteq R$ is safe or unsafe if $\mdp, r \models \varphi$ or $\mdp,r \models \neg \varphi$, respectively. Note that parameter space partitioning is also applicable if only parts of $R$ are well-defined, as well-definedness of a region is effectively decidable and such (sub-)regions can simply be tagged as \emph{not defined} and treated as being inconclusive.

As a \emph{preprocessing} step, the input model is simplified by reducing its state space.
First, \emph{bisimulation minimisation} for parametric probabilistic models~\cite{param_sttt} is used. Then, \emph{state elimination}~\cite{Daw04} is applied to all states with $V_s^\alpha = \emptyset$ and $|\Act(s)|=1$. 
We then construct the parameter-substitution of the model. As the topology of the substitution is independent of the region, for checking multiple regions we only substitute the probabilities according to the region of interest.

Now, using a heuristic from the parameter space partitioning framework, we determine a candidate region. A naive heuristic would be to start with $R$, and split the region along each dimension if no conclusive answer can be found~\cite{DBLP:conf/nfm/HahnHZ11}. More evolved heuristics apply some instantiations of the model to construct candidate regions~\cite{dehnert-et-al-cav-2015}.

%
%
For a candidate region $r \subseteq R$, regional model checking (Sections~\ref{sec:regionsMC} and~\ref{sec:nondet}) determines it to be safe or unsafe. 
Moreover, the result for a region may be \emph{inconclusive}, which 
might occur if $r$ is neither safe nor unsafe, but also if the approximation was too coarse.
The procedure stops as soon as a sufficiently large area of the parameter space $R$ has been classified into safe and unsafe regions.
%

  \section{Experimental Evaluation}
\label{sec:experiments}

We implemented and analysed the \emph{parameter lifting algorithm} (PLA) as described in Sections \ref{sec:regionsMC} and \ref{sec:nondet}. Moreover, we connected the implementation with the parameter synthesis framework \prophesy~\cite{dehnert-et-al-cav-2015}.\smallskip

\noindent\emph{Setup.}
We implemented PLA in C++. Solving the resulting non-parametric systems is done via value iteration (using sparse matrices) with a precision of $\varepsilon = 10^{-6}$.
We evaluated the performance and compared it to parameter space partitioning in \param and in \prism, both based on~\cite{DBLP:conf/nfm/HahnHZ11} and using an unsound heuristic in the implementation.
The experiments were conducted on an HP BL685C G7, 48 cores, 2.0GHz each, and 192GB of RAM. We restricted the RAM to 30GB and set a time-out of one hour for all experiments. Our PLA implementation used a single core only.
We consider the well-known pMC and pMDP benchmarks from \param's \cite{param_website} website. We additionally translated existing MDPs for a semi-autonomous vehicle~\cite{junges-et-al-tacas-2016} and the zeroconf protocol \cite{KNPS06} into pMDPs \footnote{The considered input models, properties and log files of the tools can be downloaded at \href{https://moves.rwth-aachen.de/wp-content/uploads/conference\_material/pla\_atva16.tar.gz}{moves.rwth-aachen.de/wp-content/uploads/conference\_material/pla\_atva16.tar.gz}}. 
For each instance, we analysed the parameter space $R=[10^{-5}, 1{-}10^{-5}]^{\text{\#pars}}$ until 95\% (as in \cite{DBLP:conf/nfm/HahnHZ11}) is classified as safe or unsafe. Regions for which no decisive result was found were split into equally large regions, thus mimicking the behaviour of \cite{DBLP:conf/nfm/HahnHZ11}. 
We also compared PLA to the SMT-based synthesis for pMCs in~\cite{dehnert-et-al-cav-2015}. However, using naive heuristics for determining region candidates, the SMT solver often spent too much time for checking certain regions. For the desired coverage of 95\%, this led to timeouts for all tested benchmarks.\smallskip

\noindent\emph{Results.}
The results are summarised in  Tab.~\ref{tab:experimental_results}, listing the benchmark set and the particular instance.
Further columns reflect whether a reachability or an expected reward property was checked ($\varphi$) and the number of \emph{parameters}, \emph{states} and \emph{transitions}.
We used properties as given, \eg, at the \param or \prism webpages, with a simple threshold ensuring that safe and unsafe regions exist. 
Additional information for the tested models is given in Appendix~\ref{sec:appendix:BenchmarkInformation}. 
We ran PLA in two different settings: With strong bisimulation minimisation (\emph{bisim}) and  without \emph{(direct)}.
We list the number of considered \emph{regions}, \ie, those required to cover $>95\%$ of the parameter space, and the required run time in seconds for the complete verification task, including model building and preprocessing. For \prism, we give the fastest run time producing a correct result out of 30 different possible configurations, differing in the performed bisimulation minimisation (none, strong, weak), how inconclusive regions are split (along all or along longest edge), and the order of states (all except ``random''). The \prism implementation was superior to the \param implementation in all cases. The sound variant of \prism and \param would require SMT calls similar to \cite{dehnert-et-al-cav-2015}, decreasing their performance.
{
\setlength{\tabcolsep}{4pt}
\begin{table}[t]
\smallskip\par
\centering
\scalebox{0.8}{
\begin{tabular}{cccccrrrrrrr}
	\hline
  &  &  &  &  & &  
  & \multicolumn{3}{c}{\tool{PLA}}  
  & \multicolumn{1}{c}{\tool{PRISM}} 
   \\
  \cmidrule(lr){8-10}\cmidrule(lr){11-11}
  & benchmark
  & instance 
  & $\varphi$
  & \#pars
  & \multicolumn{1}{c}{\#states} 
  & \multicolumn{1}{c}{\#trans} 
  & \multicolumn{1}{c}{\#regions} 
  & \multicolumn{1}{c}{direct} 
  & \multicolumn{1}{c}{bisim} 
  & \multicolumn{1}{c}{best}  
  \\
\hline
\hline
  \parbox[t]{3mm}{\multirow{13}{*}{\rotatebox[origin=c]{90}{\textbf{pMC}}}}
  & \multirow{8}{*}{brp} 
  &     (256,5)     & $\mathbb{P}$ & 2 &     19\,720 &      26\,627 &        37 &     \textbf{6} &    14 & TO\\
  & &    (4096,5)     & $\mathbb{P}$ & 2 &    315\,400 &     425\,987 &        13 &   \textbf{233} &    TO & TO\\
  & &     (256,5)     & $\mathbb{E}$ & 2 &     20\,744 &      27\,651 &       195 &     \textbf{8} &    15 & TO\\
  & &    (4096,5)     & $\mathbb{E}$ & 2 &    331\,784 &     442\,371 &       195 &   502 &   \textbf{417} & TO\\
  & &     (16,5)      & $\mathbb{E}$ & 4 &      1\,304 &       1\,731 & 1\,251\,220 & 2\,764 & \textbf{1\,597} & TO\\
  & &     (32,5)      & $\mathbb{E}$ & 4 &      2\,600 &       3\,459 & 1\,031\,893 &    TO & \textbf{2\,722} & TO\\
  & &     (256,5)     & $\mathbb{E}$ & 4 &     20\,744 &      27\,651 &    --     &    TO &             TO & TO\\
\cline{2-11}
  & \multirow{3}{*}{crowds} 
  &     (10,5)      & $\mathbb{P}$ & 2 &    104\,512 &     246\,082 &       123 &    17 &     \textbf{6} & 2038\\
  & &     (15,7)      & $\mathbb{P}$ & 2 &  8\,364\,409 &  25\,108\,729 &       116 & 1\,880 &   \textbf{518} & TO\\
  & &     (20,7)      & $\mathbb{P}$ & 2 & 45\,421\,597 & 164\,432\,797 &       119 &    TO & \textbf{2\,935} & TO\\
\cline{2-11}
  & \multirow{2}{*}{nand}   
  &     (10,5)      & $\mathbb{P}$ & 2 &     35\,112 &       52\,647 &       469 &    \textbf{22} &    30 & TO\tablefootnote{The fastest \prism configuration gave an incorrect answer.\label{prism:incorrect}}\\
  & &     (25,5)      & $\mathbb{P}$ & 2 &    865\,592 &   1\,347\,047 &       360 &   \textbf{735} & 2\,061 & TO\\ 
\hline
\hline
  \parbox[t]{3mm}{\multirow{12}{*}{\rotatebox[origin=c]{90}{\textbf{pMDP}}}}
  & \multirow{2}{*}{brp}  
  &     (256,5)     & $\mathbb{P}$ & 2 &     40\,721 &      55\,143 &        37 &    \textbf{35} & 3\,359 & TO\\
  & &    (4096,5)     & $\mathbb{P}$ & 2 &    647\,441 &     876\,903 &        13 & \textbf{3\,424} &    TO & TO\\
\cline{2-11}
  & \multirow{4}{*}{consensus} 
  &      (2,2)      & $\mathbb{P}$ & 2 &        272 &         492 &       119 &  $<$\textbf{1}  &  $<$\textbf{1} & 31\footnotemark[\getrefnumber{prism:incorrect}]\\
  & &     (2,32)      & $\mathbb{P}$ & 2 &      4\,112 &       7\,692 &     108 &   \textbf{113}  &   141  & TO\footnotemark[\getrefnumber{prism:incorrect}]\\
  & &      (4,2)      & $\mathbb{P}$ & 4 &     22\,656 &      75\,232 &  6\,125 & \textbf{1\,866} & 2\,022 & TO\footnotemark[\getrefnumber{prism:incorrect}]\\
  & &      (4,4)      & $\mathbb{P}$ & 4 &     43\,136 &     144\,352 &    --   &    TO           &    TO  & TO\footnotemark[\getrefnumber{prism:incorrect}]\\ 
\cline{2-11}
  & \multirow{4}{*}{sav}  
  &    (6,2,2)   & $\mathbb{P}$ & 2 &        379 &       1\,127 &       162 &  $<$\textbf{1} &  $<$\textbf{1} & TO\footnotemark[\getrefnumber{prism:incorrect}]\\
  & & (100,10,10) & $\mathbb{P}$ & 2 &  1\,307\,395 &   6\,474\,535 &        37 & \textbf{1\,612} &    TO & TO\\
  & &    (6,2,2)    & $\mathbb{P}$ & 4 &        379 &       1\,127 &   621\,175 &   944 &   \textbf{917} & TO\footnotemark[\getrefnumber{prism:incorrect}]\\
  & &   (10,3,3)   & $\mathbb{P}$ & 4 &      1\,850 &       6\,561 &           &    TO &    TO & TO\footnotemark[\getrefnumber{prism:incorrect}]\\ 
\cline{2-11}
  & \multirow{2}{*}{zeroconf}  
  &       (2)       & $\mathbb{P}$ & 2 &     88\,858 &     203\,550 &       186 &    \textbf{86} & 1\,295 & TO\\
  & &       (5)       & $\mathbb{P}$ & 2 &    494\,930 &   1\,133\,781 &       403 & \textbf{2\,400} &    TO & TO\\ 
\hline& 
\end{tabular}
}
\vspace{-2mm}
\caption{Runtimes of synthesis on different benchmark models.}
\label{tab:experimental_results}
\vspace{-5mm}
\end{table}
}

{
\setlength{\tabcolsep}{4pt}
\begin{table}[t]
\smallskip\par
\centering
\scalebox{0.8}{
\begin{tabular}{cccccrrrrrrrr}
	\hline
	& & instance
	&  $\varphi$ 
	& \#pars 
	& \multicolumn{1}{c}{\#states}
	& \multicolumn{1}{c}{\#trans}
	& \multicolumn{1}{c}{\#par trans} 
	& \multicolumn{1}{r}{$t$} 
	& \multicolumn{1}{c}{safe} 
	& \multicolumn{1}{c}{unsafe} 
	& \multicolumn{1}{c}{neither} 
	& \multicolumn{1}{c}{unkn} 
	\\ \hline
\parbox[t]{3mm}{\multirow{4}{*}{\rotatebox[origin=c]{90}{\textbf{pMC}}}} 
&	  brp    & (256,5) & $\mathbb{E}$ & 2 &  20\,744 &  27\,651   & 13\,814  &  \bf{51} & 14.9\% & 79.2\% &  \bf{5.8}\% & \bf{0.2}\% \\
&	         & (256,5) & $\mathbb{E}$ & 4 &  20\,744 &  27\,651   & 13\,814  &  \bf{71} &  7.5\% & 51.0\% & \bf{40.6}\% & \bf{0.8}\% \\
&	 crowds  & (10,5)  & $\mathbb{P}$ & 2 & 104\,512 & 246\,082   & 51\,480  &  \bf{44} & 54.4\% & 41.1\% &  \bf{4.2}\% & \bf{0.3}\% \\
&	  nand   & (10,5)  & $\mathbb{P}$ & 2 &  35\,112 &  52\,647   & 25\,370  &  \bf{21} & 21.4\% & 68.5\% &  \bf{6.9}\% & \bf{3.2}\% \\ \hline \hline
\parbox[t]{3mm}{\multirow{4}{*}{\rotatebox[origin=c]{90}{\textbf{pMDP}}}} 
&	  brp    & (256,5) & $\mathbb{P}$ & 2 &  40\,721 & 55\,143  & 27\,800& \bf{153} &  6.6\% & 90.4\% &  \bf{3.0}\% & \bf{0.0}\% \\
& consensus   &  (4,2)  & $\mathbb{P}$ & 4 &  22\,656 & 75\,232  & 29\,376& \bf{357} &  2.6\% & 87.0\% & \bf{10.4}\% & \bf{0.0}\% \\
&	sav & (6,2,2) & $\mathbb{P}$ & 4 &     379 &  1\,127   & 552&   \bf{2} & 44.0\% & 15.4\% 	 & \bf{35.4}\% & \bf{5.3}\% \\
&	zeroconf &   (2)   & $\mathbb{P}$ & 2 &  88\,858 & 203\,550 & 80\,088& \bf{186} & 16.6\% & 77.3\% &  \bf{5.6}\% & \bf{0.5}\% \\ \hline
\end{tabular}
}
\vspace{0.2cm}
\caption{Results for classification of a constant number of regions.}
\label{tab:results_ConstNumOfRegions}
\vspace{-0.9cm}
\end{table}
}

To evaluate the approximation quality, we additionally ran PLA for 625 equally large regions that were not refined in the case of indecisive results. We depict detailed results for a selection in Tab.~\ref{tab:results_ConstNumOfRegions}, where we denote model, instance, property type, number of parameters, states and transitions as in Tab.~\ref{tab:results_ConstNumOfRegions}. Column \emph{\#par trans} lists the number of transitions labeled with a non-constant function. Running times are given in column $t$. Next, we show the percentage of regions that our approach could conclusively identify as safe or unsafe. For the remaining regions, we sampled the model at the corner points to analyse the approximation error. Column \emph{neither} gives the percentage of regions for which the property is neither always satisfied, nor always violated (as obtained from the sampling). In these cases, the inconclusive result is not caused by the approximation error but by the region selection. Finally, the fraction of the remaining regions for which it is still unknown if they are safe, unsafe or neither is given in column \emph{unkn}.\smallskip

\noindent\emph{Observations.}
PLA outperforms existing approaches by several orders of magnitude. We see two major reasons. First, the approach exploits the structure of parametric models, in which transition probabilities are usually described by simple functions. This is a major benefit over state-elimination based approaches where any structure is lost. 
Secondly, the approach benefits from the speed of the numerical approaches used in non-parametric probabilistic verification. However, it is well known that problems due to numerical instability are an issue here.
Furthermore, when checking a single region, the number of parameters has only a minor influence on the runtime; more important is the number of states and the graph-structure. However, the number of required regions grows exponentially in the number of parameters. Therefore, investigating good heuristics for the selection of candidate regions proves to be essential. Nevertheless, already the naive version used here yields a superior performance.

Tab.~\ref{tab:results_ConstNumOfRegions} shows that the over-approximation of PLA is sufficiently tight to immediately cover large parts of the parameter space. In particular, for all benchmark models with two parameters, we can categorise more than 94\% of the parameter space as safe/unsafe within less than four minutes. For four parameters, we cannot cover as much space due to the poor choice of regions: A lot of regions cannot be proven (un)safe, because they are in fact neither (completely) safe nor unsafe and not because of the approximation. This is tightly linked with the observed increase in runtime for models with four parameters in Tab.~\ref{tab:experimental_results} since it implies that regions have to be split considerably before a decision can be made.
The minimal number of regions depends only on the property and the threshold used, as in \cite{DBLP:conf/nfm/HahnHZ11} and in \cite{dehnert-et-al-cav-2015}. PLA might need additional regions (although empirically, this is not significant), this corresponds to the practical case in \cite{dehnert-et-al-cav-2015} when regions are split just due to a time-out of the SMT-solver. 
\vspace{-0.3cm}

  \section{Conclusion}
\label{sec:Conclusion}
This paper presented parameter lifting, a new approach for parameter synthesis of Markov models.
It relies on replacing parameters by nondeterminism, scales well, and naturally extends to treating parametric MDPs.

  \bibliographystyle{splncs}
  \bibliography{literature}
  
  \clearpage
  \appendix
  \section{Proofs}
\begin{proof}[of Theorem~\ref{th:mdpApprox} on Page~\pageref{th:mdpApprox}]
We prove the second statement.
A proof for the first statement can then be derived in a straightforward manner.
$\pmdp, r \models \neg \reachProplT$ holds iff for all $u \in r$ there is a scheduler $\sched \in \Sched{\pmdp}$ for which the reachability probability in the MC $\pmdp^\sched[u]$ exceeds the threshold $\lambda$, \ie,
\[
\pmdp, r \models \neg \reachProplT \iff \min_{u\in r} \max_{\sched \in \Sched{\pmdp}}  \reachPrT[{\pmdp^\sched[u]}]  > \lambda.
\]
A lower bound for this probability is obtained as follows:
\begin{align*}
    &\min_{u\in r}\max_{\sched \in \Sched{\pmdp}} \big( \reachPrT[{\pmdp^\sched[u]}] \big)
\ge \max_{\sched \in \Sched{\pmdp}} \min_{u\in r}\big( \reachPrT[{\pmdp^\sched[u]}] \big) \overset{\ast}{\ge}\\
 &\max_{\sched \in \Sched{\pmdp}} \min_{\altsched \in \Sched{\subst{\pmdp^\sigma}{r}}}\big( \reachPrT[(\subst{\pmdp^\sigma}{r})^\altsched] \big) 
\overset{\ast\ast}{=}   \max_{\sched \in \Sched[\pOne]{\sg}}  \min_{\altsched \in \Sched[\pTwo]{\sg}} \big(\reachPrT[\sg^{\sched, \altsched}] \big)\end{align*}
The inequality $\ast$ is due to Corollary~\ref{cor:dtmcApprox}, where $\sg =\subst{\pmdp}{r}$.
The equality $\ast\ast$ holds by Corollary~\ref{cor:mdpApprox}.
Then:
\[
\sg \models_{\{\pOne\}} \p_{> \lambda}(\finally T) \iff \max_{\sched \in \Sched[\pOne]{\sg}}  \Big( \min_{\altsched \in \Sched[\pTwo]{\sg}} \big(\reachPrT[\sg^{\sched, \altsched}] \big) \Big) > \lambda.
\]
\end{proof}
  \section{Additional Benchmark Information}
\label{sec:appendix:BenchmarkInformation}
Tab.~\ref{tab:model_information} gives additional information for the benchmarks presented in Section~\ref{sec:experiments}.
In addition to the values in Tab.~\ref{tab:experimental_results}, we depict the number of states and transitions of the reduced model (applying strong bisimulation) in the columns labeled by \emph{bisimulation}.
Moreover, columns \emph{save} and \emph{unsave} indicate the fractions of the parameter space which have been identified as save and unsafe, respectively.
{
\setlength{\tabcolsep}{4pt}
\begin{table}[t]
\smallskip\par
\centering
\scalebox{0.8}{
\begin{tabular}{cccccrrrrrrr}
	\hline
  &  &  &  & 
  & \multicolumn{2}{c}{{unreduced}}
  & \multicolumn{2}{c}{{bisimulation}}
  & \multicolumn{2}{c}{{fractions}}
   \\
  \cmidrule(lr){6-7} \cmidrule(lr){8-9} \cmidrule(lr){10-11}
  & benchmark
  & instance 
  & $\varphi$
  & \#pars
  & \multicolumn{1}{c}{\#states} 
  & \multicolumn{1}{c}{\#trans} 
  & \multicolumn{1}{c}{\#states} 
  & \multicolumn{1}{c}{\#trans} 
  & \multicolumn{1}{c}{\%save} 
  & \multicolumn{1}{c}{\%unsafe}  
  \\
\hline
\hline
  \parbox[t]{3mm}{\multirow{13}{*}{\rotatebox[origin=c]{90}{\textbf{pMC}}}}
  & \multirow{8}{*}{brp} 
  &     (256,5)     & $\mathbb{P}$ & 2 &     19\,720 &      26\,627 &     10503    &   14855             &  5,5\%   &	89,8 \% \\
  & &    (4096,5)     & $\mathbb{P}$ & 2 &    315\,400 &     425\,987 &     \multicolumn{2}{c}{{TO}}     &  1,6\%   &   93,8\%     \\
  & &     (256,5)     & $\mathbb{E}$ & 2 &     20\,744 &      27\,651 &      11531  &   15883            &  15,0\%   &   80,0\%     \\
  & &    (4096,5)     & $\mathbb{E}$ & 2 &    331\,784 &     442\,371 &     184331   &  253963           &  15,0\%   &    80,0\%     \\
  & &     (16,5)      & $\mathbb{E}$ & 4 &      1\,304 &       1\,731 &  731& 1003                       &  27,5\%   &   67,6\%     \\
  & &     (32,5)      & $\mathbb{E}$ & 4 &      2\,600 &       3\,459 & 1451 &   1995                    &  23,2\%   &    71,8\%     \\
  & &     (256,5)     & $\mathbb{E}$ & 4 &     20\,744 &      27\,651 &   11531   & 15883                &  \multicolumn{2}{c}{{TO}} \\
\cline{2-11}
  & \multirow{3}{*}{crowds} 
  &     (10,5)      & $\mathbb{P}$ & 2 &    104\,512 &     246\,082 &      80  &   120                   &  53,7\%   &  41,3\%         \\
  & &     (15,7)      & $\mathbb{P}$ & 2 &  8\,364\,409 &  25\,108\,729 &    120    & 180                &   41,1\% &	53,9\%   \\
  & &     (20,7)      & $\mathbb{P}$ & 2 & 45\,421\,597 & 164\,432\,797 &     120   &  180               &   41,6\%   &  53,4\%  \\
\cline{2-11}
  & \multirow{2}{*}{nand}   
  &     (10,5)      & $\mathbb{P}$ & 2 &     35\,112 &       52\,647 &     23602   &  34092              & 21,8\%   &  73,3\%      \\
  & &     (25,5)      & $\mathbb{P}$ & 2 &    865\,592 &   1\,347\,047 &    673115    &  1030424         & 20,6\%   &   74,4\%     \\ 
\hline
\hline
  \parbox[t]{3mm}{\multirow{12}{*}{\rotatebox[origin=c]{90}{\textbf{pMDP}}}}
  & \multirow{2}{*}{brp}  
  &     (256,5)     & $\mathbb{P}$ & 2 &     40\,721 &      55\,143 &    21032     &  29750              &  5,5\%   &   89,8\%     \\
  & &    (4096,5)     & $\mathbb{P}$ & 2 &    647\,441 &     876\,903 &        \multicolumn{2}{c}{{TO}}  &  1,6\%   &  93,8\%     \\
\cline{2-11}
  & \multirow{4}{*}{consensus} 
  &      (2,2)      & $\mathbb{P}$ & 2 &        272 &         492 &     153   &  332                     & 28,0\%   &  66,99      \\
  & &     (2,32)      & $\mathbb{P}$ & 2 &      4\,112 &       7\,692 &   2793   & 6092                  & 25,0\%   & 70,02      \\
  & &      (4,2)      & $\mathbb{P}$ & 4 &     22\,656 &      75\,232 &  9351 & 37609                    &  6,3\%   &  88,8\%     \\
  & &      (4,4)      & $\mathbb{P}$ & 4 &     43\,136 &     144\,352 &   19271   &    77545             &    \multicolumn{2}{c}{{TO}}    \\ 
\cline{2-11}
  & \multirow{4}{*}{sav}  
  &    (6,2,2)   & $\mathbb{P}$ & 2 &        379 &       1\,127 &    166   & 776                         & 51,3\%   &  43,8\%      \\
  & & (100,10,10) & $\mathbb{P}$ & 2 &  1\,307\,395 &   6\,474\,535 &    \multicolumn{2}{c}{{TO}}        & 6,3\%   &  89,1\%      \\
  & &    (6,2,2)    & $\mathbb{P}$ & 4 &        379 &       1\,127 &  166 &   776                        & 64,1\%   &  30,9\%      \\
  & &   (10,3,3)   & $\mathbb{P}$ & 4 &      1\,850 &       6\,561 &   950        &   4940               &    \multicolumn{2}{c}{{TO}}\\ 
\cline{2-11}
  & \multirow{2}{*}{zeroconf}  
  &       (2)       & $\mathbb{P}$ & 2 &     88\,858 &     203\,550 &     26423   &  67056               &  17,4\%   &  77,6\%     \\
  & &       (5)       & $\mathbb{P}$ & 2 &    494\,930 &   1\,133\,781 &    \multicolumn{2}{c}{{TO}}     &  38,2\%   &   56,9\%     \\ 
\hline& 
\end{tabular}
}
\vspace{-2mm}
\caption{Additional information for the considered benchmark instances.}
\label{tab:model_information}
\vspace{-5mm}
\end{table}
}

\end{document}